\documentclass[aps,prd,preprint,floatfix,superscriptaddress]{revtex4}

\usepackage{graphicx}

\def\ra{\rightarrow}

\def\bwt{\begin{widetext}}
\def\ewt{\end{widetext}}
\def\be{\begin{equation}}
\def\ee{\end{equation}}
\def\bea{\begin{eqnarray}}
\def\eea{\end{eqnarray}}
\def\bean{\begin{eqnarray*}}
\def\eean{\end{eqnarray*}}
\def\bary{\begin{array}}
\def\eary{\end{array}}
\def\bit{\begin{itemize}}
\def\eit{\end{itemize}}

\def\ra{\rightarrow}

\begin{document}

\setcounter{page}{0}
\thispagestyle{empty}

\preprint{ANL-HEP-PR-02-057, hep-ph/0210135}
\bigskip

\title{Differential Cross Section for Higgs Boson Production 
Including All-Orders Soft Gluon Resummation}

\author{Edmond~L.~Berger}
\email[e-mail: ]{berger@anl.gov}
\affiliation{High Energy Physics Division, 
Argonne National Laboratory, Argonne, IL 60439}
\author{Jianwei~Qiu}
\email[e-mail: ]{jwq@iastate.edu}
\affiliation{Department of Physics and Astronomy, 
Iowa State University, Ames, IA 50011}

\date{\today}

\begin{abstract}
The transverse momentum $Q_T$ distribution is computed for inclusive 
Higgs boson production at the energy of the CERN Large Hadron Collider.  
We focus on the dominant gluon-gluon subprocess in perturbative 
quantum chromodynamics and incorporate contributions 
from the quark-gluon and quark-antiquark channels.  Using an impact-parameter 
$b$-space formalism, we include all-orders resummation of large logarithms 
associated with emission of soft gluons.  Our resummed results merge 
smoothly at large $Q_T$ with the fixed-order expectations in perturbative
quantum chromodynamics, as they should, with no need for a matching 
procedure.  They show a high degree of stability with respect to 
variation of parameters associated with the non-perturbative input at 
low $Q_T$.  We provide distributions $d\sigma/dy dQ_T$ for Higgs boson 
masses from $M_Z$ to 200~GeV.  The average transverse momentum at zero 
rapidity $y$ grows approximately linearly with mass of the Higgs boson over 
the range $M_Z < m_h <200$~GeV, $<Q_T> \simeq 0.18 m_h + 18 $~GeV.  We provide 
analogous results for $Z$ boson production, for which we compute 
$<Q_T> \simeq 25$~GeV.  The harder transverse momentum distribution for the 
Higgs boson arises because there is more soft gluon radiation in Higgs boson 
production than in $Z$ production.  
\end{abstract}


\maketitle


\section{INTRODUCTION}

The breaking of electroweak symmetry in the standard model (SM) of 
elementary particle interactions is achieved through the Higgs 
mechanism.  In the simplest realization, a complex Higgs doublet is 
introduced, and a single neutral CP-even Higgs boson is predicted.    
Two Higgs doublets are expected at low energies in supersymmetric 
extensions of the standard model.  The lightest neutral CP-even 
Higgs particle $h$ resembles the SM Higgs boson in most regions of 
parameter space of broken supersymmetry~\cite{Gunion:1984yn}.
Direct experimental searches at the CERN Large Electron Positron 
Collider (LEP) based on the presumption of significant decay 
branching ratio into bottom ($b$) quarks place the mass of a SM-like 
Higgs state above approximately 115 GeV~\cite{lephiggs}. An 
alternative analysis, based on the assumption of Higgs boson decay 
into hadronic jets, without $b$-tagging, provides a bound of about 
113 GeV~\cite{flavorind}.  A good theoretical description of electroweak 
data~\cite{Abbaneo:2001ix} in the framework of the SM requires the 
Higgs boson to be lighter than about 200 GeV~\cite{lephiggs}.  Within 
the minimal supersymmetric standard model (MSSM), the upper bound 
on the mass of the lightest Higgs state is roughly 
135 GeV~\cite{Heinemeyer:1998jw}.  The search for the Higgs boson is 
a central motivation for the experimental programs at the Fermilab 
Tevatron and the CERN Large Hadron Collider (LHC), with detection 
techniques guided by theoretical expectations about its production 
dynamics and decay properties. 

A Higgs boson resembling one anticipated by the SM is expected to be 
discovered at the LHC through various partonic production processes 
and decays to SM particles~\cite{Cavalli:2002vs,atlas:1999fr}.  These include
\begin{itemize}
 \item $g g \ra h X$, with $h \ra \gamma \gamma$, $h \ra W^+ W^-$, or $h \ra Z Z$ ;
 \item $g g \rightarrow t \bar{t} h X$, with $h \ra b \bar{b}$ or $h \ra \gamma \gamma$ ; and 
 \item $W^+ W^- (ZZ) \ra h X$, with $h \ra W^+ W^-$, $h \ra \gamma \gamma$, or 
       $h \ra \tau^+ \tau^-$.
\end{itemize}

The fully inclusive gluon-gluon fusion 
subprocess~\cite{Wilczek:1977zn,Graudenz:1992pv} 
$g g \ra h X$ is the dominant mechanism of production.  
In this process, production occurs through triangle loops of colored (s)particles 
that couple to the Higgs boson 
and to gluons.  In the SM, the most relevant contribution is from a loop of 
top ($t$) quarks, with a small contribution from bottom quark loops.  In 
supersymmetric theories, the bottom quark loop contribution can be important 
for small $m_A$ and large values of $\tan\beta$, where $m_A$ is the mass of 
the pseudo-scalar Higgs boson and $\tan\beta$ is the ratio of Higgs vacuum 
expectation values. In the MSSM, loops of bottom squarks may make an 
important contribution if the mass of the bottom squark is light 
enough~\cite{Berger:2002vs}. 

Semi-inclusive production of the Higgs boson in association with top quarks, 
$g g \rightarrow t \bar{t} h X$, 
has a relatively low rate because of the large masses in the final state. 
The semi-inclusive weak boson fusion modes~\cite{Kauer:2000hi} 
$W^+ W^- (ZZ) \ra h X$ require identification of (forward) ``trigger'' 
jets from subprocesses such as 
$q q \rightarrow q q W W \rightarrow q q h X$.  
Searches at the Tevatron rely on associated production of the Higgs 
boson with weak bosons, $W$ and $Z$, and the decay mode 
$h \ra b \bar b$~\cite{Carena:2000yx}. 

For masses $m_h$ of the Higgs boson up to about 150 GeV, the SM Higgs boson produced 
in the gluon fusion process should be discovered at the LHC in the rare 
di-photon decay mode $h \rightarrow \gamma \gamma$~\cite{Cavalli:2002vs,atlas:1999fr}.  
Once data are in hand, the immense $\gamma \gamma$ background from 
various sources can be determined from the data by a measurement of the 
di-photon invariant mass distribution $d \sigma/d M_{\gamma \gamma}$ on 
both sides of the Higgs resonance.  Beforehand, precise theoretical 
calculations of the expected differential cross sections for production of 
the signal and backgrounds~\cite{Berger:1983yi,Balazs:1997hv,Bern:2002jx} 
are important for quantitative 
evaluation of the required measurement accuracies and detector performance.  
Good estimations of the expected transverse transverse momentum distributions 
can suggest selections in this variable that should improve background 
rejection.  In this paper, we concentrate on behavior of the Higgs boson 
transverse momentum distribution in the region of small and intermediate 
values of $Q_T$.  

Above $W^+ W^-$ threshold but below $Z$-pair threshold, the decay mode 
$h \rightarrow W^+ W^-$ has a branching ratio approaching 100\%.  
Because there are missing neutrinos in the leptonic decay modes of 
the $W$'s, the invariant mass distribution of the Higgs boson cannot be 
reconstructed directly, but the transverse mass spectrum of the 
($\ell^+$ $\ell^-$ $\not E_T$) system can be measured.  This distribution 
is expected to be broad, and straightforward sideband determinations 
of the background are not possible.  However, measurements of the 
backgrounds in regions in which the signal is expected to be absent can 
then be extrapolated into the expected signal region provided that good 
predictions are available of differential cross sections.  

In the SM, the coupling of gluons to the Higgs boson through the top quark 
loop is simplified in the limit of large top quark mass $m_t$.  In this limit, 
the relevant Feynman diagrams are obtained from an effective 
Lagrangian~\cite{Ellis:1975ap}.  For inclusive production integrated over 
all transverse momentum, the next-to-leading order contributions in 
perturbative quantum chromodynamics (QCD) are reported in 
Ref.~\cite{Dawson:1990zj}. The $m_t \rightarrow \infty$ approximation is 
valid to an accuracy of $\sim 5$\% for $m_h \le 2 m_t$.  
Within this approach, the total cross section 
for $g g \rightarrow h X$, is known to next-to-next-to-leading order 
accuracy~\cite{Harlander:2002wh}. 

Computations of the transverse momentum distribution for Higgs boson 
production at leading order in perturbative QCD~\cite{Ellis:1987xu} show that 
the large $m_t$ approximation serves well when the Higgs boson mass $m_h$ is 
less than twice the top quark mass and the Higgs boson transverse momentum 
$Q_T$ is less than $m_t$.  The next-to-leading order contributions to the 
transverse momentum distribution are computed with numerical integration 
methods in Ref.~\cite{deFlorian:1999zd}. The doubly-differential cross section 
in rapidity and transverse momentum of the Higgs boson is computed analytically 
at next-to-leading order in Refs.~\cite{Ravindran:2002dc}~and~\cite{Glosser:2002gm}.  

When the transverse momentum $Q_T$ is comparable to the mass $m_h$ of the 
Higgs boson, there is only one hard momentum scale in the perturbative 
expansion of the cross section as a function of the strong coupling 
$\alpha_s$, and fixed-order computations in perturbative QCD are expected to 
be applicable. However, in the region $Q_T \ll m_h$, where the cross section 
is greatest, the coefficients of the expansion in $\alpha_s$ depend 
functionally on logarithms of the ratio of the two quantities, 
$m_h$ and $Q_T$.  A well-studied case is the cross section for gauge boson 
(W, Z, or virtual photon) production in hadronic collisions, at measured 
transverse momentum.  The logarithmic term is 
$(\alpha_s/\pi) \ln^2(Q^2/Q_T^2)$ where $Q$ is the mass of the boson and 
$Q_T$ its transverse momentum.  For Higgs boson production at $m_h = 125$~GeV, 
the peak of the transverse momentum distribution is expected to occur near 
$Q_T = 14$~GeV at LHC energies, as is shown in this paper.  Correspondingly, 
$\ln^2(m_h^2/Q_T^2) \sim 19$ and $(\alpha_s(\mu)/\pi) \ln^2(m_h^2/Q^2_T) \sim 0.7$ 
if $\mu = m_h$ or $\sim 1.1$ if $\mu = Q_T$.  In either case, the relevant 
expansion parameter in the perturbative series is close to 1, meaning that 
higher order contributions in perturbation theory are not suppressed.  When 
$Q_T \ll m_h$, straightforward 
perturbation theory is inapplicable. Indeed, even at leading order in 
$\alpha_s$, the computed cross section $d\sigma/dQ_T^2$ includes a term 
proportional to $(\alpha_s/Q_T^2) \ln (m_h^2/Q_T^2)$.  This divergence as 
$Q_T \rightarrow 0$ is obviously unphysical.  

All-orders resummation is the established method for mastering the large 
logarithmic coefficients of the expansion in $\alpha_s$ and for obtaining 
well-behaved cross sections at intermediate and 
small $Q_T$~\cite{Collins:1981uk,Collins:1984kg,Ellis:1997ii}.  It is a 
reorganization of the perturbative expansion that brings the large 
logarithmic terms under systematic control, providing finite predictions
for the $Q_T$ dependence that are different from those at fixed order.  
Resummation overcomes the limitations of fixed-order calculations but 
still incorporates all the information that these calculations offer. 

Resummation may be carried out in either $Q_T$~\cite{Ellis:1997ii} or 
impact parameter ($b$) 
space, which is the Fourier conjugate of $Q_T$ space.  All else being 
equal, the $b$ space approach has the advantage that transverse 
momentum conservation is explicit.  Using renormalization 
group techniques, Collins, Soper, and Sterman (CSS)~\cite{Collins:1984kg} 
devised a $b$ space 
resummation formalism that resums all logarithmic terms as singular 
as $(1/Q_T^2)\ln^n(Q^2/Q_T^2)$ when $Q_T \rightarrow 0$.  This 
formalism has been 
used widely for computations of the transverse momentum distributions of 
vector bosons in hadron 
reactions~\cite{Davies:1984sp,Arnold:1990yk,Ladinsky:1993zn,Balazs:1995nz,Balazs:1997xd1,Berger:1998ev,Qiu:2000hf,Landry:1999an,Zhang:2002yz}, 
Higgs bosons~\cite{Catani:vd,Hinchliffe:1988ap,Kauffman:1991jt,Yuan:1991we,deFlorian:2000pr}, 
and other processes~\cite{Balazs:1997hv,Berger:yp}.  

In many of these applications of the CSS formalism, two difficulties  
significantly limit its predictive power in the region of small 
and intermediate values of $Q_T$.  The first of these is the 
ambiguity, if not discontinuity in $Q_T$, associated with the ``matching'' 
of the 
resummed and fixed-order calculations in a region of $Q_T$ below which the 
resummed form is used, and above which a fixed-order expression is used.  
The second difficulty is associated with the quantitative importance of 
assumed forms for non-perturbative functions that, in many implementations, 
dominate the $Q_T$ distribution in the region of very small $Q_T$ and 
affect even the $Q_T$ distribution at large $Q_T$.  In recent 
papers~\cite{Qiu:2000hf}, Qiu and Zhang demonstrate that both of these 
drawbacks can be overcome.  By using an integral form for the oscillatory 
Bessel function in the Fourier transform, they show that the $b$-space 
resummation procedure produces smooth distributions for all $Q_T < Q$, 
obviating the need for a matching prescription.  Second, to substantially 
reduce the influence of phenomenological non-perturbative functions, they 
derive a new functional expression that extrapolates the perturbatively 
calculated $b$-space resummation expression from the region of small $b$,  
where it is valid, to the region of large $b$ where non-perturbative 
input is needed.  Their analysis shows that perturbation theory 
itself indicates the form of the required non-perturbative power 
corrections.  They also demonstrate that the predictive power of 
resummation improves considerably with total center-of-mass collision 
energy $\sqrt{S}$.  In this paper, we further develop the methodology of 
Qiu and Zhang and use it to study the production of Higgs bosons.   

In Sec.~II, we review the pertinent aspects of the $b$-space resummation 
formalism, with particular emphasis on Higgs boson production.  Working at 
next-to-leading-logarithm (NLL) accuracy, we resum both leading and 
next-to-leading logarithmic terms associated with soft gluon emission to 
all orders in $\alpha_s$.  At LHC energies, the typical values of the incident 
parton momentum fractions $x_A\sim x_B\sim m_h/\sqrt{S} \sim 0.009$ (for 
$m_h = 125$~GeV) are small, and the gluon distribution evolves steeply at 
small $x$. Consequently, the saddle point in $b$ of the Fourier transform 
from $b$-space to $Q_T$ space is well into the region of perturbative 
validity.  Resummed perturbation theory is valid at small $b$, just as 
perturbative QCD is valid at large $Q_T$.  However, the Fourier transform 
from $b$-space to $Q_T$-space requires specification of the $b$-space 
distribution for all $b$.  In Sec.~III, we show that the 
region of large $b$ becomes progressively less important quantitatively as 
$\sqrt{S}$ increases.  The $Q_T$ distribution at LHC energies is entirely 
insensitive to the functional form we employ for the non-perturbative input 
at large $b$, so long as the form used for this extrapolation does not 
modify the $b$-space distribution at small $b$.  In Sec.~III.B, we discuss 
the extrapolation from the region of small $b$ to the region of large $b$, 
and we specify the form we use for the non-perturbative functions at large 
$b$. We discuss why other approaches in the literature may introduce 
inappropriate dependence on $\sqrt{S}$.  
We provide analytic expressions for the fixed-order perturbative 
subprocesses in Sec.~IV.  We employ expressions for the parton-level 
hard-scattering functions valid through first-order in $\alpha_s$, 
including contributions from the glue-glue, quark-glue, and quark-antiquark 
incident partonic subprocesses. 

Numerical predictions are presented in Sec.~V. 
We show the doubly differential cross section $d\sigma/dy dQ_T$ as a 
function of $Q_T$ at fixed rapidity $y = 0$, and the $Q_T$ 
distributions for rapidity integrated over the interval $|y| < 2.4$.  
We present results for four choices of mass of the Higgs boson, values that 
span the range of present interest in the SM, from $m_h = M_Z = 91.187$~GeV 
to $m_h = 200$~GeV.  To illustrate interesting differences, we also show our 
results for $Z$ boson production at the same energy.  Our resummed results 
make a smooth transition to the fixed-order perturbative results near or just 
above $Q_T = Q$, for all $Q$, without need of a supplementary matching 
procedure.  Our implementation of the resummation formalism works smoothly 
throughout the $Q_T$ region even for the distributions differential in rapidity, 
$d\sigma/dy dQ_T$.  Two points are evident in the comparison of $Z$ boson and 
Higgs boson production, with $m_h = M_Z$.  The peak in the $Q_T$ distribution 
occurs at a smaller value of $Q_T$ for $Z$ production 
(4.8~GeV {\em vs.} 10~GeV) and the distribution is 
narrower for $Z$ production.  The larger QCD color factors produce more gluonic 
showering in the glue-glue scattering subprocess that dominates inclusive Higgs 
boson production than in the fermionic subprocesses relevant for $Z$ production.  
After all-orders resummation, the enhanced showering suppresses the large-$b$ 
(small $Q_T$) region more effectively for Higgs boson production.  We compare 
the predicted $Q_T$ distributions for Higgs boson production at different 
masses. The peak of the distribution shifts to greater $Q_T$ as $m_h$ grows, in
approximately linear fashion, and the distribution broadens somewhat.  
The mean value $<Q_T>$ grows from about 35 GeV at $m_h = M_Z$ to about 54 GeV 
at $m_h = 200$ GeV, and the root-mean-square grows from about 59~GeV to about 
87~GeV.  For $Z$ production, we find $<Q_T> = 25$~GeV and 
$<Q_T^2>^{1/2} = 38$~GeV.  The harder $Q_T$ spectrum suggests that the signal to 
background ratio can be enhanced if Higgs bosons are selected with large $Q_T$.  

Choices of variable parameters are made in obtaining our results, and
we examine the sensitivity of the results to these choices, including 
the renormalization/factorization scale $\mu$ and the non-perturbative 
input.  Scale dependence is the most important source of uncertainty. 
It can shift the position of the peak by about 1~GeV, with corresponding 
changes in the normalization of the distribution above and below the 
position of the peak.  The value of $d\sigma/dydQ_T$ at the peak position 
is shifted by 4 to 5\%.  Changes in the parameters of the non-perturbative 
input produce effects that at most 1 to 2\% depending on the size 
of the power corrections we introduce.  In the formulation we use to describe 
the non-perturbative region, there is essentially no effect on the behavior of 
differential cross section at large $Q_T$.  In comparison with prior work, we 
note that the locations of the maxima in the distributions $d\sigma/dydQ_T$ 
occur at slightly larger values of $Q_T$ in our case, and the distributions 
themselves differ as a function of $Q_T$ above the location of the maximum.  
The differences arise from the different treatment of the non-perturbative 
input.  In our approach, the assumed parametrization of non-perturbative 
effects has the desirable property that it does not affect the physics in the 
perturbative region $b < 0.5$~GeV.  

Conclusions are summarized in Sec.~VI.
%
\section{All-orders Resummed $Q_T$ Distribution}

We consider the inclusive hadronic reaction in which a color neutral heavy 
boson of invariant mass $Q$ is produced: 

\begin{equation}
A(P_A) + B(P_B)\rightarrow C(Q) + X ,
\end{equation}
with $C=\gamma^*, W^\pm, Z$, or a  Higgs boson in the limit in which 
the top quark mass $m_t\gg Q/2$.  The square of the total center-of-mass 
energy of the collision is $S$.  At the LHC, $\sqrt{S} = 14$~TeV.  
In the CSS resummation formalism, the differential cross section is 
written as the sum 

\begin{equation}
\frac{d\sigma_{AB\rightarrow C X}}{dQ^2\, dy\, dQ_T^2}
=
\frac{d\sigma_{AB\rightarrow C X}^{\rm (resum)}}{dQ^2\, dy\, dQ_T^2}
+
\frac{d\sigma_{AB\rightarrow C X}^{\rm (Y)}}{dQ^2\, dy\, dQ_T^2}\, .
\label{css-gen}
\end{equation}

\noindent
The all-orders resummed term is a Fourier transform from $b$-space

\begin{eqnarray}
\frac{d\sigma_{AB\rightarrow C X}^{\rm (resum)}}{dQ^2\, dy\, dQ_T^2}
&=&
\frac{1}{(2\pi)^2}\int d^2b\, e^{i\vec{Q}_T\cdot \vec{b}}\,
W_{AB\rightarrow C X}(b,Q,x_A,x_B)
\label{css-resum}
\nonumber \\
&=& 
\int \frac{db}{2\pi}\, J_0(Q_T\, b)\, 
bW_{AB\rightarrow C X}(b,Q,x_A,x_B), 
\end{eqnarray}
where $J_0$ is a Bessel function. The function 
$W_{AB\rightarrow C X}(b,Q,x_A,x_B)$
resums to all orders in QCD perturbation theory the singular terms that 
would otherwise behave as $\delta^2(Q_T)$ and $(1/Q_T^2)\ln^m(Q^2/Q_T^2)$, 
for all $m \ge 0$. The variables $x_A$ and $x_B$ are 
light-cone momentum fractions carried by the incident partons from 
hadrons $A$ and $B$: 
\begin{equation}
x_A= \frac{Q}{\sqrt{S}}\, e^y 
\quad\mbox{and}\quad
x_B= \frac{Q}{\sqrt{S}}\, e^{-y} , 
\end{equation}
and $y$ is the rapidity of the heavy boson.  The variables $x_A$ and 
$x_B$ do not depend on $Q_T$.  

Resummation treats only the parts of the fixed-order QCD expression that 
are at least as singular as $Q_T^{-2}$ in the limit $Q_T \rightarrow 0$.  
The remainder, including possible less singular pieces of the fixed-order 
perturbative contribution, is defined as the difference of the cross 
section computed at fixed order $n$ in perturbation theory and its 
$Q_T \ll Q$ asymptote that is at least as singular as $Q_T^{-2}$.     
\begin{equation}
\frac{d\sigma_{AB\rightarrow C X}^{\rm (Y)}}{dQ^2\, dy\, dQ_T^2}\, 
=
\frac{d\sigma_{AB\rightarrow C X}^{\rm (pert)}}{dQ^2\, dy\, dQ_T^2}
-
\frac{d\sigma_{AB\rightarrow C X}^{\rm (asym)}}{dQ^2\, dy\, dQ_T^2}\, .  
\label{css-Y}
\end{equation}
This remainder is not significant quantitatively at modest $Q_T$, since 
the dominant singularities of the two terms on the right-hand-side cancel 
in the region $Q_T \rightarrow 0$.  However, the difference 
becomes important when $Q_T \sim Q$.  Explicit expressions for the 
fixed-order remainder terms are presented in Sec.~IV.  

We may factor out the lowest order partonic cross section and rewrite 
the function $W_{AB\rightarrow C X}(b,Q,x_A,x_B)$ that appears in the 
integrand of Eq.~(\ref{css-resum}) as 

\begin{equation}
W_{AB\rightarrow C X}(b,Q,x_A,x_B) = \sum_{ij}
W_{ij}(b,Q,x_A,x_B) \, \sigma^{(0)}_{ij\rightarrow C X}(Q) .
\end{equation}
Here, $\sigma^{(0)}_{ij\rightarrow C X}(Q)$ is the 
lowest order cross section for a pair of partons of flavor $i$ and $j$
to produce a heavy boson of invariant mass $Q$.
Because the heavy boson is color neutral, the parton
flavors $ij$ can be either the quark-antiquark or the gluon-gluon
combination.  

\subsection {Higgs boson production}

Specializing to Higgs boson $h$ production in gluon-gluon fusion (and 
using $m_h$ and $Q$ interchangeably for the Higgs boson mass 
in the remainder of this paper), we write 
\begin{equation}
W_{AB\rightarrow h X}(b,Q,x_A,x_B) =
W_{gg}(b,Q,x_A,x_B) \, \sigma^{(0)}_{gg\rightarrow h X}(Q) .
\end{equation}
In the effective Lagrangian approximation~\cite{Ellis:1975ap,Dawson:1990zj} 
where we keep only the 
contribution from the top quark loop, the lowest order partonic cross 
section for $gg \rightarrow h X$ is

\begin{equation}
\sigma_{gg\rightarrow h X}^{(0)}(Q)
=\sigma_0\, \frac{\pi}{S}\, m_h^2\, \delta(Q^2-m_h^2) ,
\label{gg-H-lo}
\end{equation}
with
\begin{equation}
\sigma_0 = \left(\sqrt{2}G_F\right) 
           \frac{\alpha_s^2(\mu_r)}{576\pi} .
\label{sigma-0}
\end{equation}
In this expression, $G_F$ is the Fermi constant, and $\mu_r$ denotes a 
renormalization scale.  We note that $\sigma_0$ is formally of second 
order in the strong coupling strength. However, we {\em define} $\sigma_0$ 
be the effective 0'th order cross section and focus on the influence of 
yet higher order contributions.  If we were to include contributions other 
than from top quarks in the loop, whether from bottom quarks or bottom 
squarks, we could easily replace $\sigma_0$ by a different expression.  This  
inclusion would change the overall normalization of our predictions.  For 
on-mass-shell gluons, QCD corrections to the $g g h$ effective vertex are 
known to order $\alpha^2_s$~\cite{Chetyrkin:1997iv} and likewise could be 
incorporated in a change in the numerical value of the effective $\sigma_0$.  

For comparison with Eq.~(\ref{gg-H-lo}), we also present the expression 
for the lowest order massive 
lepton-pair Drell-Yan cross section:

\begin{equation}
\sigma_{q\bar{q}\rightarrow \ell^+\ell^- X}^{(0)}(Q)
= e_q^2 \, \frac{4\pi^2\alpha_{em}^2}{9SQ^2} .  
\label{drellyan-lo}
\end{equation}


\subsection {Evolution equation for resummation}

Resummation of the large logarithmic terms is achieved in the CSS formalism 
through a solution of the evolution equation 
\begin{equation}
\frac{\partial}{\partial\ln Q^2} W_{gg}(b,Q,x_A,x_B)
= \left[ K_g(b\mu,\alpha_s(\mu)) + G_g(\frac{Q}{\mu},\alpha_s(\mu)) \right]
  W_{gg}(b,Q,x_A,x_B)\, ,
\label{css-W-evo}
\end{equation}
where the subscripts $g$ and $gg$ indicate the gluonic process.
There are two physical momentum variables in this expression, $Q$ and 
$1/b$, as well dependence on the scale variable $\mu$ that, in 
principle, could take on values from of order $1/b$ to of order $Q$.    
However, the $\mu$ dependence of $K_g$ and $G_g$ is determined by 
renormalization group equations (RGE's),
\begin{eqnarray}
\frac{\partial}{\partial\ln\mu^2} K_g(b\mu,\alpha_s(\mu))
&=& -\frac{1}{2} \gamma_g(\alpha_s(\mu)) \, ,
\label{css-K-rg} \\
\frac{\partial}{\partial\ln\mu^2} G_g(\frac{Q}{\mu},\alpha_s(\mu))
&=& \frac{1}{2} \gamma_g(\alpha_s(\mu)) \, .
\label{css-G-rg}
\end{eqnarray}
\noindent
The RGE's for $K_g$ and $G_g$ ensure that 
$d/d\ln(\mu^2)[W_{gg}(b,Q,x_A,x_B)] = 0$ perturbatively, an expression that  
is consistent with the fact that $W_{gg}$ is effectively a physical observable.
The coefficients $\gamma_g^{(n)}$ in an expansion of the anomalous dimension 
$\gamma_g(\alpha_s(\mu))=\sum_{n=1} \gamma_g^{(n)} (\alpha_s(\mu)/\pi)^n$ 
are calculable perturbatively.  

The RGE's for $K_g$ and $G_g$ may be integrated over $\ln(\mu^2)$ 
from $\ln(C_1^2/b^2)$ to $\ln(\mu^2)$, 
and from $\ln(\mu^2)$ to $\ln(C_2^2Q^2)$, respectively, where 
$C_1$ and $C_2$ are integration constants.  
The values of $C_1$ and $C_2$ are somewhat arbitrary. 
Following CSS, we select 
$C_1 = c = 2e^{-\gamma_E}={\cal O}(1)$, where Euler's constant 
$\gamma_E\approx 0.577$, and $C_2 = 1$.   After the integrations, 
one derives 
\begin{equation}
K_g(b\mu,\alpha_s(\mu)) + G_g(Q/\mu,\alpha_s(\mu)) 
= -\int_{c^2/b^2}^{Q^2}\, 
  \frac{d\bar{\mu}^2}{\bar{\mu}^2}\, 
  A_g(\alpha_s(\bar{\mu}))
- B_g(\alpha_s(Q))\, ,
\label{css-KG}
\end{equation}
where $A_g$ is a function of $\gamma_g(\alpha_s(\bar{\mu}))$ and
$K_g(c,\alpha_s(\bar{\mu}))$ while $B_g$ depends on both
$K_g(c,\alpha_s(Q))$ and $G_g(1,\alpha_s(Q))$. The choice of
$C_1 = c = 2e^{-\gamma_E}$ and $C_2 = 1$ has the virtue 
of removing logarithmic dependence on these parameters from the 
functions $A_g$ and $B_g$.   
 
The functions $A_g$ and $B_g$ are free of logarithmic 
dependence and have well-behaved perturbative expansions 

\begin{equation}
A_g(\alpha_s(\bar{\mu})) =
\sum_{n=1} A_g^{(n)}\left(\frac{\alpha_s(\bar{\mu})}{\pi}\right)^n
\quad\mbox{and}\quad
B_g(\alpha_s(\bar{\mu})) =
\sum_{n=1} B_g^{(n)}\left(\frac{\alpha_s(\bar{\mu})}{\pi}\right)^n  .
\label{AB-def}
\end{equation}

\noindent 
The first two coefficients in the expansions for $A$ and $B$ are 
known~\cite{Kauffman:1991jt,Yuan:1991we,deFlorian:2000pr}.  For the 
choices $C_1 = c = 2e^{-\gamma_E}$ and $C_2 = 1$, these are 
\begin{eqnarray}
A_g^{(1)} &=& C_A\ ,
\nonumber\\
A_g^{(2)} &=& \frac{C_A}{2} \left[
           C_A\left(\frac{67}{18}-\frac{\pi^2}{6}\right)
           -\frac{5}{9}\, n_F \right]\, ,
\nonumber\\
B_g^{(1)} &=& -\frac{11C_A-2 n_F}{6}\, ,
\nonumber\\
B_g^{(2)} &=& C_A^2 
            \left(\frac{23}{24}+\frac{11}{18}\pi^2
                 -\frac{3}{2}\zeta(3)\right)
          + \frac{1}{2}\,C_F\, n_F
\nonumber \\
        &-& C_A\, n_F \left(\frac{1}{12}+\frac{\pi^2}{9}\right)
          - \frac{11}{8}\, C_A\, C_F .
\label{css-AB}
\end{eqnarray}
Here, $\zeta(3)$ is the third Riemann function, $n_F = 5$ is the
number of active quark flavors, and  
\begin{equation}
C_A = N_c = 3,
\quad\mbox{and}\quad
C_F=\frac{N_c^2-1}{2N_c} = \frac{4}{3}
\end{equation}
for SU(3) color. The corresponding fermionic coefficients $A_q^{(i)}$ and 
$B_q^{(i)}$, with $i = 1,2$ are found in Refs.~\cite{Collins:1984kg,Davies:1984sp}. 
The coefficient $A_q^{(3)}$ is available in numerical form~\cite{Vogt:2000ci}.   

The solution of the linear evolution Eq.~(\ref{css-W-evo}) may be 
written in the form 
\begin{equation}
W_{gg}(b,Q,x_A,x_B) = {\rm e}^{-S_g(b,Q)}\,  
W_{gg}(b,\frac{c}{b},x_A,x_B) .
\end{equation}
The function $W_{gg}(b,c/b,x_A,x_B)$ depends on only 
one momentum variable, $1/b$, and it can be computed in perturbation theory 
as long as $1/b$ is large enough.  All large logarithmic terms from 
$\ln(c^2/b^2)$ to $\ln(Q^2)$ are resummed to all orders in $\alpha_s$ in the 
exponential factor, with 
\begin{equation}
S_g(b,Q) = \int_{c^2/b^2}^{Q^2}\, 
  \frac{d\bar{\mu}^2}{\bar{\mu}^2} \left[
  \ln\left(\frac{Q^2}{\bar{\mu}^2}\right) 
     A_g(\alpha_s(\bar{\mu})) 
   + B_g(\alpha_s(\bar{\mu})) \right]\, .
\label{css-S}
\end{equation}

\noindent
The appearance of $(c/b)^2$ as the lower limit of integration serves to 
cut off the non-perturbative region of small $Q_T$. 

The function $W_{gg}(b,c/b,x_A,x_B)$ may be written 
in factored form as~\cite{Collins:1984kg,Ellis:1997ii} 

\begin{equation}
W_{gg}(b,\frac{c}{b},x_A,x_B) 
= f_{g/A}(x_A,\mu,\frac{c}{b})\, f_{g/B}(x_B,\mu,\frac{c}{b}) ,
\label{css-W-cb}
\end{equation}

\noindent
where $f_{g/A}$ and $f_{g/B}$ are modified gluon parton distributions

\begin{equation}
f_{g/A}(x_A,\mu,\frac{c}{b}) = \sum_a 
  \int_{x_A}^1\frac{d\xi}{\xi}\, \phi_{a/A}(\xi,\mu)\,
  C_{a\rightarrow g}\left(\frac{x_A}{\xi},\mu,\frac{c}{b} \right) .
\label{mod-pdf}
\end{equation}
Here we see the normal appearance of a factorization scale $\mu$, the 
scale of factorization between short-distance hard-scattering and the 
long-distance physics represented by the parton densities.  The sum 
in Eq.~(\ref{mod-pdf}) runs over all parton types, and $\phi_{a/A}(\xi,\mu)$ 
is the usual momentum ($\xi$) distribution of parton $a$ from hadron 
$A$ at the scale $\mu$.  The short-distance coefficient functions are 
expanded in a perturbative series as 
\begin{equation}
C_{a\rightarrow b}\left(z,\mu,\frac{c}{b} \right)
=\sum_{n=0} 
C_{a\rightarrow b}^{(n)}(z,\mu,b)\left(\frac{\alpha_s(\mu)}{\pi}\right)^n .
\end{equation}
The first two coefficients in the expansion of $C$ are 
published~\cite{Kauffman:1991jt,Yuan:1991we,deFlorian:2000pr}: 
\begin{eqnarray}
C^{(0)}_{g\rightarrow g}(z,\mu,\frac{c}{b}) &=& \delta(1-z)\, ,
\nonumber\\
C^{(0)}_{i\rightarrow g}(z,\mu,\frac{c}{b}) &=& 0\, ,
\nonumber \\
C^{(1)}_{g\rightarrow g}(z,\mu,\frac{c}{b}) &=& \delta(1-z)
  \left[ C_A \left(\frac{\pi^2}{4} + \frac{5}{4}\right) 
        -\frac{3}{4}C_F
  \right] 
\nonumber \\
&-& P_{g\rightarrow g}(z)\,
     \ln\left(\frac{\mu b}{c}\right) , 
\nonumber \\
C^{(1)}_{i\rightarrow g}(z,\mu,c/b) &=& \frac{1}{2}C_F\, z
  - P_{i\rightarrow g}(z)\,
     \ln\left(\frac{\mu b}{c}\right) .
\label{css-coef}
\end{eqnarray}
Here $i$ represents a quark or antiquark flavor, and 
$P_{g\rightarrow g}(z)$ and $P_{i\rightarrow g}(z)$ are leading 
order parton-to-parton splitting functions,
\begin{eqnarray}
P_{g\rightarrow g}(z)
&=&
  2C_A\left[\frac{z}{(1-z)_+}+\frac{(1-z)}{z}+z(1-z)\right] 
 +\frac{11C_A-2n_F}{6}\,\delta(1-z) , 
\nonumber \\
P_{i\rightarrow g}(z)
&=&
  C_F \left[\frac{1+(1-z)^2}{z}\right] . 
\end{eqnarray}


There is only one momentum scale, $1/b$, involved in the hard-scattering 
function $W_{gg}(b,c/b,x_A,x_B)$.  Therefore, it is natural to choose 
$\mu=c/b$ in the modified parton densities $f_g(x,\mu,c/b)$ in order 
to remove the logarithmic terms in Eq.~(\ref{css-coef}).  However, this 
choice is not required.   In principle, $\mu$ may be any large momentum 
scale, {\it e.g.}, $Q$.  This freedom is equivalent to the statement  
$d/d\ln(\mu^2)[W_{gg}(b,c/b,x_A,x_B)] =0$ order-by-order in powers of 
$\alpha_s$; or,  $d/d\ln\mu^2 f_{g}(x_A,\mu,c/b) = 0$ order-by-order in 
$\alpha_s$ for $f_{g}$ defined in Eq.~(\ref{mod-pdf}).  The $\mu$-dependence 
in $C_{a\rightarrow g}$ is compensated by the $\mu$-dependence in the 
parton densities.  To order $\alpha_s$, we derive 
\begin{eqnarray}
\frac{d}{d\ln\mu^2}f_{g}(x,\mu,\frac{c}{b})
&=& 
\sum_{a} \int_{x}^1 \frac{d\xi}{\xi}\left[
\frac{d}{d\ln\mu^2}\phi_{a}(\xi,\mu)\right] 
C^{(0)}_{a\rightarrow g}(\frac{x}{\xi},\mu,\frac{c}{b})
\nonumber\\
&+&
\sum_{a} \int_{x}^1 \frac{d\xi}{\xi}
\phi_{a}(\xi,\mu) \left[
\frac{d}{d\ln\mu^2}
C^{(1)}_{a\rightarrow g}(\frac{x}{\xi},\mu,\frac{c}{b})
\right]
\nonumber \\
&=& 
\sum_{a} \int_{x}^1 \frac{d\xi}{\xi}\left[
\frac{d}{d\ln\mu^2}\phi_{a}(\xi,\mu)\right] 
\delta_{ag}\delta(1-\frac{x}{\xi})
\nonumber\\
&+&
\sum_{a} \int_{x}^1 \frac{d\xi}{\xi}
\phi_{a/A}(\xi,\mu) \left[
- \frac{\alpha_s}{2\pi} P_{a\rightarrow g}(\frac{x}{\xi})\right]
\nonumber \\
&=& 0 .
\end{eqnarray}
The last equality follows because both terms on the right-hand-side equal
$(d/d\ln\mu^2)\phi_{g}(x,\mu)$.

For simplicity, we elect to drop the subscript $AB\rightarrow h X$ 
for the $b$-space distribution $W_{AB\rightarrow h X}(b,Q,x_A,x_B)$.  
In its region of validity at small $b$ it takes the factored form 

\begin{equation}
W^{\rm pert}(b,Q,x_A,x_B)
=
\sigma_{gg\rightarrow h X}^{(0)}\sum_{a,b}
\left[\phi_{a/A}\otimes C_{a\rightarrow g}\right]
\otimes
\left[\phi_{b/B} \otimes C_{b\rightarrow g} \right]
\times {\rm e}^{-S(b,Q)} , 
\label{css-pert}
\end{equation}
where $\otimes$ represents a convolution over parton momentum fractions.
Since leading-twist perturbative QCD and normal parton densities are 
valid only for a hard scale $\mu > \mu_0 \sim 1$ to 2 GeV, we expect 
naively that $W^{\rm pert}(b,Q,x_A,x_B)$ will be reliable for 
$b < 1/\mu_0 \sim 1.0$~GeV$^{-1}$.

To perform the Fourier transform in Eq.~(\ref{css-resum}), we require an  
expression for $W(b,Q,x_A,x_B)$ that can be used over the entire range of 
integration in $b$. To the extent that the region of large $b$ is important 
in the integrand, we must devise a functional form for 
$W(b,Q,x_A,x_B)$ that extends into the region of large $b$, a topic 
addressed quantitatively in Sec.~III.~B.  

In the CSS formalism, the coefficient functions $C_{i\rightarrow j}$ and
the Sudakov form factor $S(b,Q)$ are process dependent.  As discussed 
recently in Ref.~\cite{Catani:2000vq}, it is possible to reorganize the
perturbatively calculated coefficient functions $C_{i\rightarrow j}$ 
and the Sudakov form factor $S(b,Q)$ to make them universal.  This 
universality is achieved if an all orders process-dependent hard part
$H_{gg}(\alpha_s(Q))$ is introduced into Eq.~(\ref{css-pert}),  
\begin{equation}
W^{\rm pert}_{H}(b,Q,x_A,x_B)
=
\sigma_{gg\rightarrow hX}^{(0)}\, H_{gg}(\alpha_s(Q))\, \sum_{a,b}
\left[\phi_{a/A}\otimes C_{a\rightarrow g}\right]
\otimes
\left[\phi_{b/B} \otimes C_{b\rightarrow g} \right]
\times {\rm e}^{-S(b,Q)} ,
\label{css-pert-modified}
\end{equation}
with $H_{gg}(\alpha_s(Q))=\sum_n H^{(n)}_{gg} (\alpha_s/\pi)^n$, and
the lowest order $H_{gg}^{(0)}=1$.  In analogy with parton distributions, 
the universal coefficient functions $C_{i\rightarrow j}$ and 
the Sudakov form factor $S(b,Q)$ can be defined by choosing a
``resummation scheme''.  The higher order contributions to the hard
part, $H^{(n)}_{gg}$, can be calculated perturbatively.  Since the 
reorganization affects only the perturbative part of $W(b,Q,x_A,x_B)$, 
our approach to extrapolate $W^{\rm pert}_{H}(b,Q,x_A,x_B)$ into the 
nonperturbative large $b$ region, discussed in Sec.~III.~B, could also 
be used to calculate the $Q_T$ distribution in this modification of the 
CSS formalism.  

\section{Role of the region of large impact parameter}
 
\subsection{Suppressed role of the region of large $b$ at large $\sqrt{S}$}

Prior to presenting alternative expressions for the extrapolation of 
$W(b,Q,x_A,x_B)$
into the region of large $b$, we emphasize one important property of the 
$b$-space resummation formalism: the resummed exponential factor 
$\exp[-S(b,Q)]$ suppresses the $b$-integrand in Eq.~(\ref{css-resum}) 
when $b$ is larger than $1/Q$.  Therefore, when $Q \gg \mu_0$, the 
Fourier transform is dominated by a region of $b$ much smaller than
$1/\mu_0$, and the calculated $Q_T$ distribution is insensitive to
the non-perturbative information at large $b$.  

Saddle-point methods are used in Refs.~\cite{Parisi:1979se,Collins:1984kg} 
to show that 
for large enough $Q$, the Fourier transform in Eq.~(\ref{css-resum}) is 
dominated by an impact parameter 
\begin{equation}
b_{\rm SP} \simeq \frac{1}{\Lambda_{\rm QCD}}
  \left( \frac{\Lambda_{\rm QCD}}{Q} \right)^{\lambda} , 
\label{css-saddle}
\end{equation}
where $\lambda=2A^{(1)}/(2A^{(1)} + \beta_0) \approx 0.61$ and $\approx 0.41$, 
for gluonic and fermionic processes, respectively, for $n_f=5$ quark flavors. 
This equation shows that the momentum scale corresponding to the
saddle point, $1/b_{\rm SP}$, should be within the perturbative
region if $Q$ is large enough.  In Ref.~\cite{Qiu:2000hf}, a more 
complete expression is 
derived for $b_{\rm SP}$ that takes into account the full $b$ dependence of 
$W_{ij}(b,c/b,x_A,x_B)$.  

The saddle point at $Q_T=0$ is determined from the differential equation 
\begin{equation}
 \frac{d}{db}\ln\left(b\, {\rm e}^{-S(b,Q)}\right)_{b=b_0}
+\frac{d}{db}\ln\left(\sum_{ij}\sigma_{ij\rightarrow h X}(Q)\,
{W}_{ij}(b,\frac{c}{b},x_A,x_B)\right)_{b=b_0}
=0 .
\label{saddle}
\end{equation}
If ${W}_{ij}(b,c/b,x_A,x_B)$ has a weak dependence on $b$ 
near $b_0$, the second term in Eq.~(\ref{saddle}) can be neglected,
and the first term yields $b_0\approx b_{\rm SP}$.  However, as we 
now show, the $b$-dependence of ${W}_{ij}(b,c/b,x_A,x_B)$ is 
directly proportional to the evolution of the modified parton
distributions.  

The second term on the left-hand side of Eq.~(\ref{saddle}) can be 
expressed as
\begin{equation}
\frac{d}{db} \ln \left(
\sigma_{gg}(Q) W_{gg}(b,\frac{c}{b},x_A,x_B)
\right)
=
\frac{d}{db} \left( \ln f_{g/A}(x_A,\mu,\frac{c}{b})
                    +\ln f_{g/B}(x_B,\mu,\frac{c}{b}) \right) ,
\nonumber
\end{equation}
and
\begin{eqnarray}
\frac{d}{db} f_{g/A}(x_A,\mu,\frac{c}{b})
&=& 
\frac{1}{b}
\sum_{a} \int_{x_A}^1 \frac{d\xi}{\xi}
\phi_{a/A}(\xi,\mu)\left[
\frac{d}{d \ln b}
C_{a\rightarrow g}(\frac{x_A}{\xi},\mu,\frac{c}{b})
\right]
\nonumber \\
&\approx &
\frac{1}{b}
\sum_{a} \int_{x_A}^1 \frac{d\xi}{\xi}
\phi_{a/A}(\xi,\mu)\left[
- \frac{\alpha_s}{\pi} P_{a\rightarrow g}(\frac{x_A}{\xi})\right]
\nonumber \\
&\approx &
-\frac{2}{b}
\frac{d}{d\ln\mu^2}\phi_{g/A}(x_A,\mu)\ .
\label{mu-evo}
\end{eqnarray}
\noindent 
The derivative $(d/d\ln\mu)\phi(x,\mu)$ is typically 
positive (negative) for $x<x_0\sim 0.1$
($x>x_0$), and the evolution is very steep when $x$ is far from 
$x_0$. Correspondingly, the second term in Eq.~(\ref{saddle}) should
be important when the relevant values of $x_A \sim Q/\sqrt{S}$ and 
$x_B \sim Q/\sqrt{S}$ are much smaller (or much greater) than $x_0$. 
At the LHC, $x_A \sim x_B \sim 0.009$ for $m_h = 125$~GeV.  
\begin{figure}[ht]
\centerline{\includegraphics[width=9.0cm]{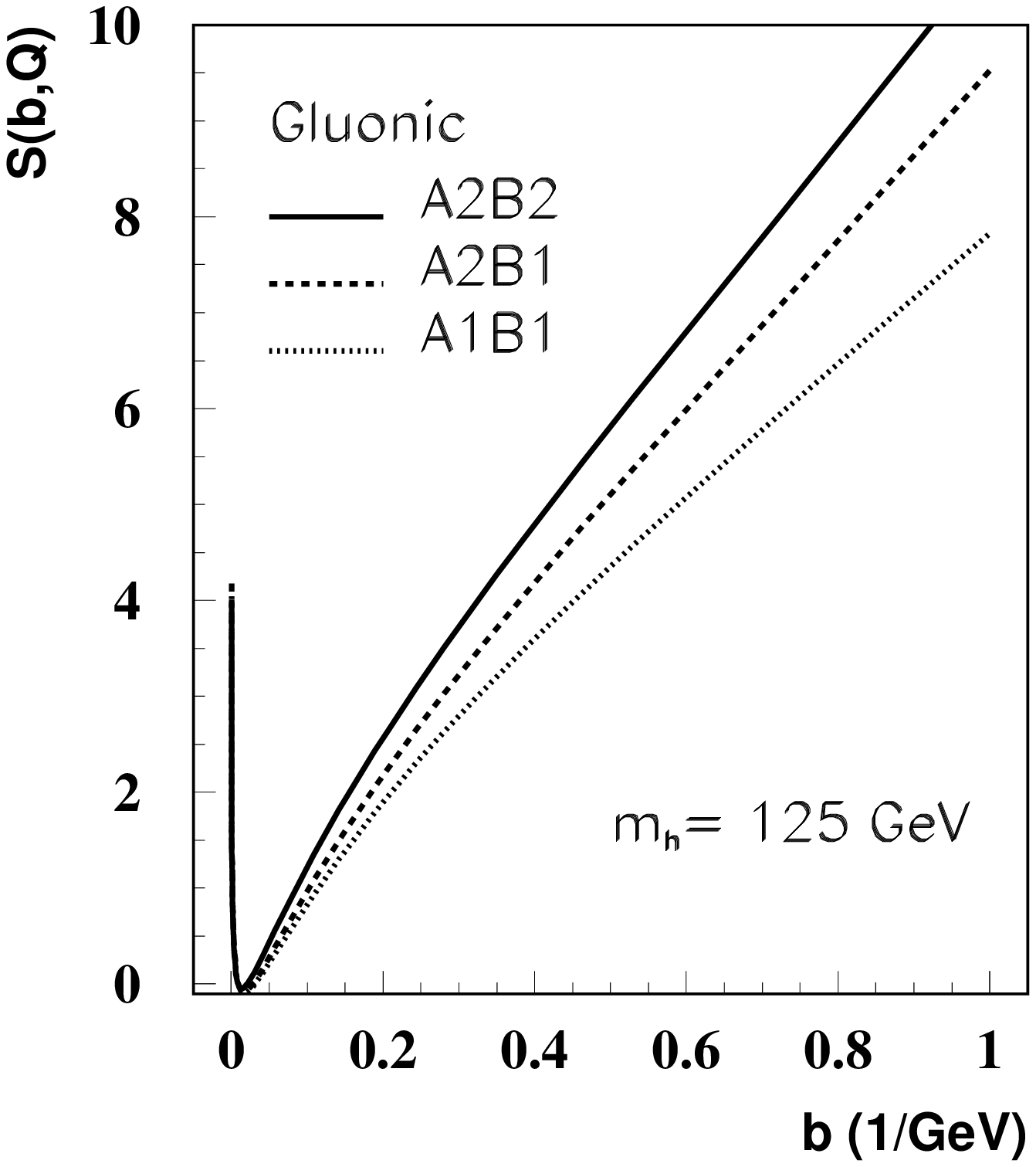}
\includegraphics[width=9.0cm]{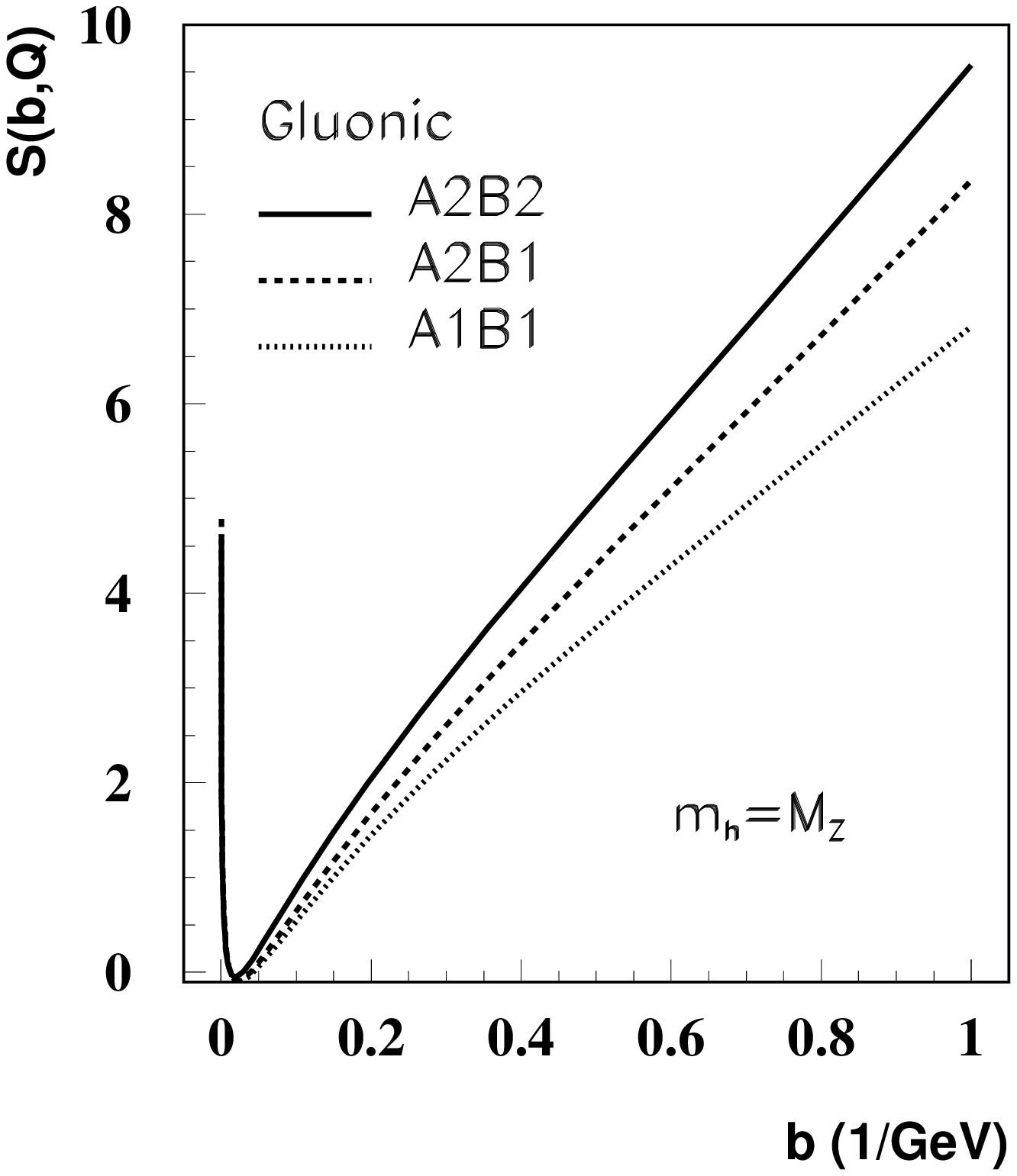}} 
\caption[]{\it The function $S(b,Q)$ for Higgs boson production in 
  gluon-gluon fusion at (a) $Q=m_h=125$ GeV and (b) $m_h = M_Z$ as a 
  function of impact parameter 
  $b$.  We show results for different orders of accuracy in the perturbative 
  expansions of the functions $A_g$ and $B_g$. The dotted curve is obtained 
  with only the first order contributions $A^{(1)}_g$ and 
  $B^{(1)}_g$ retained. For the dashed curve, we include $A^{(2)}_g$ in 
  addition, and for the solid curve, we keep all terms through 
  $A^{(2)}_g$ and $B^{(2)}_g$ in Eq.~(\ref{css-AB}). }
\label{fig:SudH}
\end{figure}

The first term alone in Eq.~(\ref{saddle}) is a decreasing function of $b$. 
The derivative vanishes at $b=b_{\rm SP}$ producing a first 
approximation for the saddle point solution, Eq.~(\ref{css-saddle}). 
The negative sign in
Eq.~(\ref{mu-evo}), and the fact that the number of small $x$ partons
increases when the scale $\mu$ increases, mean that the second
term in Eq.~(\ref{saddle}) is negative when $x_A$ and $x_B$ are
smaller than the typical $x_0$.  The full solution to Eq.~(\ref{saddle}) 
provides a saddle point that is less than the first approximation 
$b=b_{\rm SP}$. (Conversely, at fixed-target energies where the typical 
$x_A$ and $x_B$ are greater than $x_0$, the typical $b$ is greater 
than $b_{\rm SP}$.)  The numerical value for the saddle point depends 
strongly on $\sqrt{S}$, improving the predictive power of $b$-space 
resummation at large energies.
When $Q_T>0$, the Bessel function $J_0(z=Q_T b)$ further suppresses
the large $b$ region of the $b$-integration.  The suppression 
is greater for greater $Q_T$, and the $b$-space resummation formalism 
is expected to work progressively better for larger $Q_T$.  

In the remainder of this subsection, we present figures that illustrate the 
$b$ dependences of the Sudakov function $S(b,Q)$ and of the factor 
$bW(b,Q)$ in the integrand of Eq.~(\ref{css-resum}), and we comment on 
their physical implications.  The Sudakov 
function depends only on $Q$ and on the perturbatively calculable 
functions $A_g$ and $B_g$, Eqs.~(\ref{AB-def} - \ref{css-S}).  Selecting 
$Q = m_h = 125$ GeV, we show the behavior of $S(b,m_h)$ in 
Fig.~\ref{fig:SudH}. The function has a pronounced minimum at very small 
$b$, $b \sim 0.02$~GeV$^{-1}$.  Since it is $\exp[-S(b,Q)]$ that enters the 
integrand, Fig.~\ref{fig:SudH} shows that the Sudakov factor 
strongly suppresses both the large and small regions of $b$.  
It is instructive to examine the relative importance of next-to-leading 
order contributions in the functions $A_g$ and $B_g$. As shown in the 
figure, it may be sufficient at very small $b$ to retain only the first-order 
contributions, $A^{(1)}_g$ and $B^{(1)}_g$.  However, the second-order 
contributions, both $A^{(2)}_g$ and $B^{(2)}_g$, are of growing relevance 
as $b$ grows.  The larger Sudakov function at large $b$ in next-to-leading 
order means that the region of small $Q_T$ is more suppressed, resulting 
in a shift of the peak of the distribution $d\sigma/dydQ_T$ to somewhat 
larger $Q_T$ at higher orders.  The large change associated with $A^{(2)}_g$ 
may be understood because $A_g$ in Eq.~(\ref{css-S}) is multiplied by 
$\ln(Q^2)$.  It is possible that inclusion of $A^{(3)}_g$, once it is known, 
will produce important effects. A comparison of the results in 
Figs.~\ref{fig:SudH}(a)~and~(b) shows that the Sudakov function is peaked 
more narrowly at greater $m_h$.  The suppression of the region of large $b$ 
is more pronounced for larger $m_h$, again influencing a shift of the peak 
of the distribution $d\sigma/dydQ_T$ to somewhat larger $Q_T$ as $m_h$ is 
increased.  
\begin{figure}[ht]
\centerline{\includegraphics[width=9.0cm]{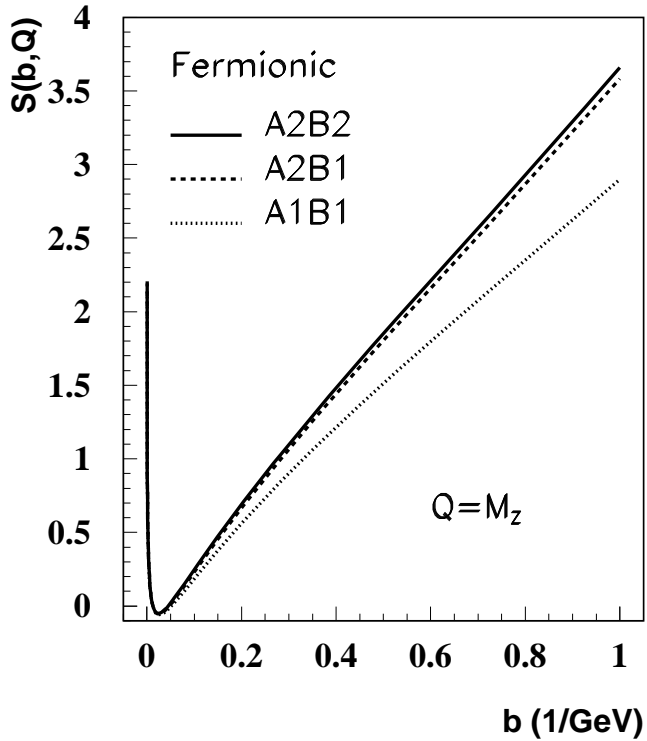}} 
\caption[]{\it The function $S(b,Q)$ for $Z$ boson production in 
  as a function of impact parameter $b$.  We show results for different 
  orders of accuracy in the perturbative 
  expansions of the functions $A_g$ and $B_g$.}
\label{fig:SudZ}
\end{figure}

To examine the differences between $Z$ production and Higgs boson production, 
we show in Fig.~\ref{fig:SudZ} the results appropriate for $Z$ boson production, 
with $Q=M_Z=91.187$ GeV.  The second-order fermionic function $B^{(2)}_q$ 
plays a much less significant role in $Z$ production than does the gluonic 
function $B^{(2)}_g$ in Higgs boson 
production.  Comparison of Figs.~\ref{fig:SudH}(b) and ~\ref{fig:SudZ} shows 
that the gluonic Sudakov function is much larger than its quark counterpart, 
but the curves are similar in shape otherwise.  At $b = 1$, the ratio is 
about 2.5.  The fact that the gluonic $A_g$ and $B_g$ functions are larger 
than their fermionic counterparts $A_q$ and $B_q$ is related to the differences 
in the values of the structure constants $C_A$ and $C_F$.  The larger color 
factor for gluons means that gluons radiate more readily.  There will be more 
gluonic showering and correspondingly more suppression of the region of large 
$b$ ({\em i.e.,} small $Q_T$) when gluonic subprocesses are dominant. Greater 
suppression of the region of large $b$ means that perturbation theory will 
have greater predictive power and resummed calculations will be more reliable 
for gluonic processes.  

\begin{figure}[ht]
\centerline{\includegraphics[width=9.0cm]{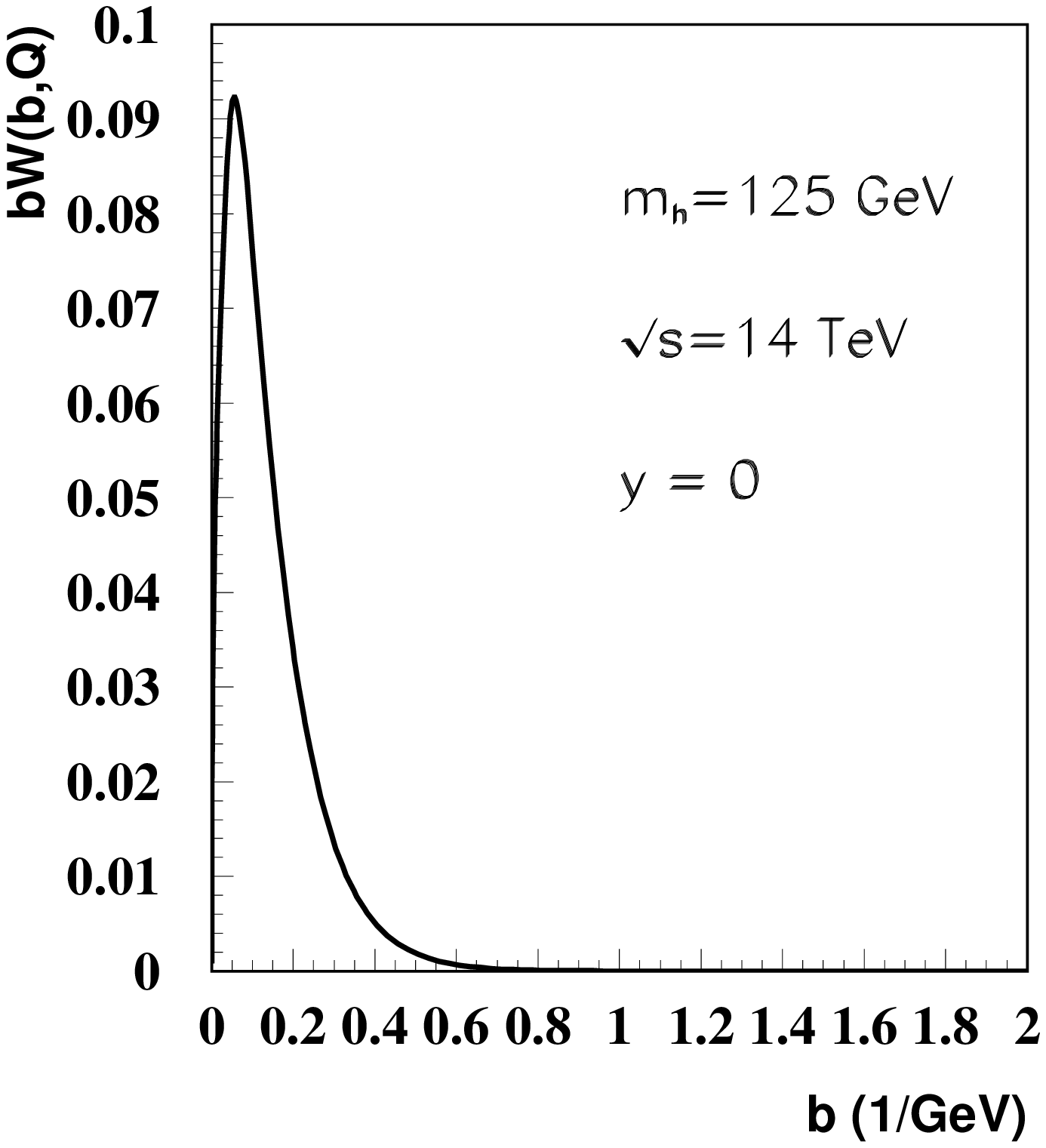} 
\includegraphics[width=9.0cm]{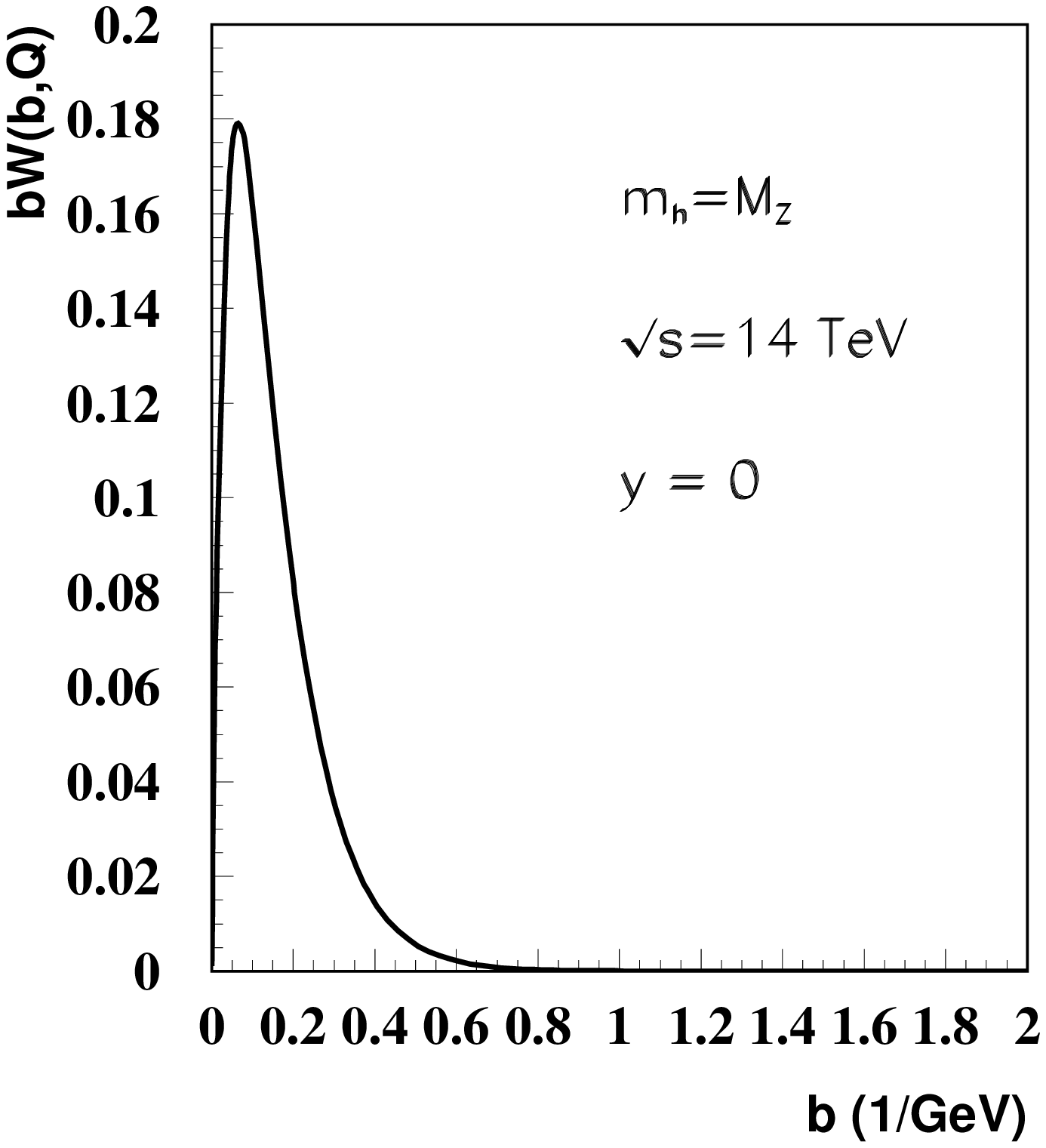}}
\caption[]{\it The all-orders $b$-space resummed function $bW(b,Q)$ for Higgs 
  boson production in gluon-gluon fusion at $\sqrt{S} = 14$~TeV for (a) 
  $Q=m_h=125$ GeV and (b) $m_h = M_Z$ as a function of impact parameter 
  $b$.} 
\label{fig:bWHLHC}
\end{figure}
\begin{figure}[ht]
\centerline{\includegraphics[width=9.0cm]{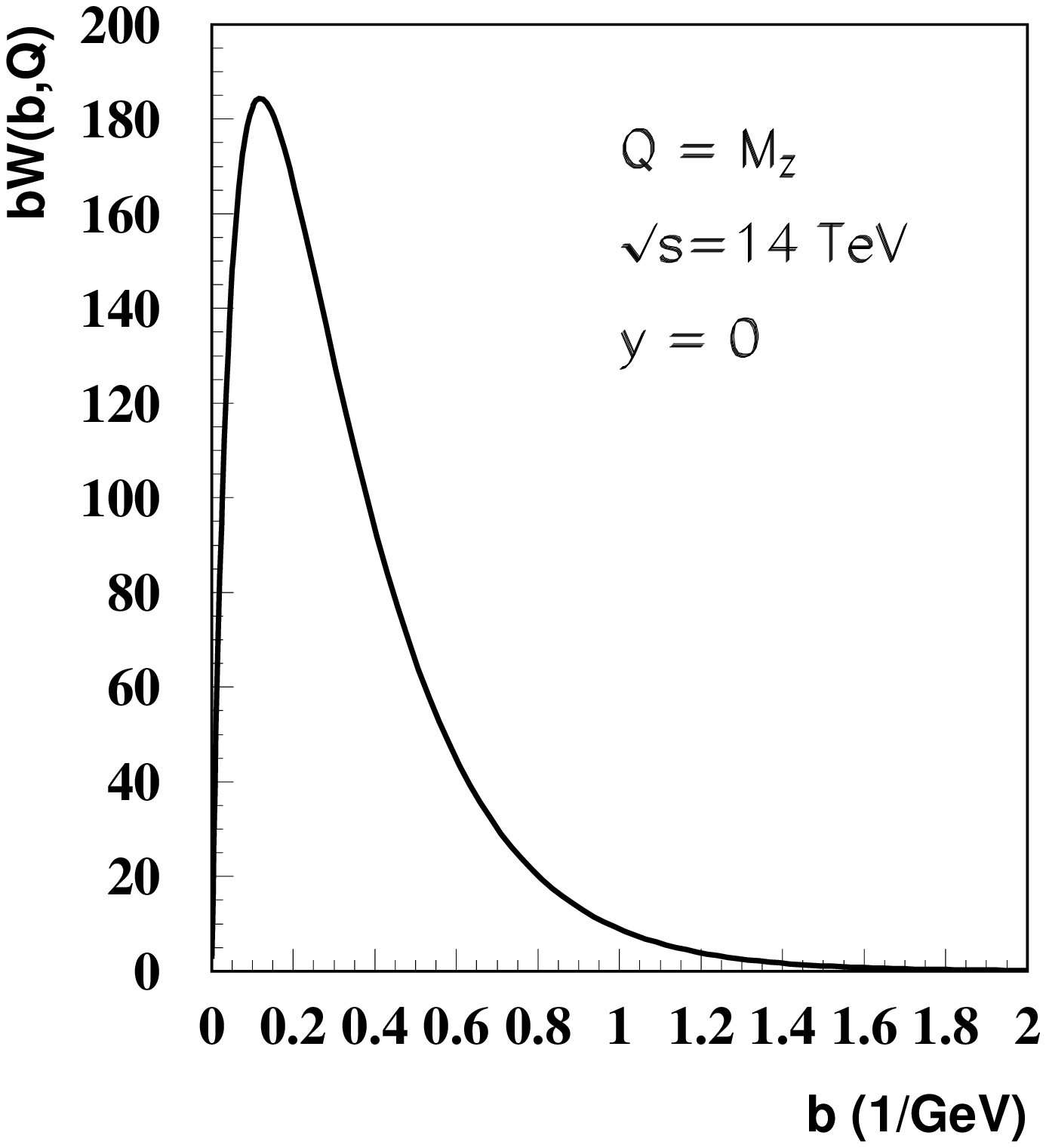}} 
\caption[]{\it The all-orders $b$-space resummed function $bW(b,Q)$ for $Z$ 
  boson production at $\sqrt{S} = 14$~TeV as a function of impact parameter 
  $b$.} 
\label{fig:bWZLHC}
\end{figure}

The all-orders resummed $b$-space function $W(b,Q)$, Eq.~(\ref{css-pert}), 
depends on the choice of parton densities.  We use the CTEQ5M 
set~\cite{Lai:1999wy}.  In Fig.~\ref{fig:bWHLHC}, we show $bW(b,Q)$ as
a function of $b$ at the energy of the LHC for two choices of the Higgs 
boson mass.  This function is peaked sharply near 
$b \sim 0.05$~GeV$^{-1}$ ({\em c.f.}, $Q_T \sim 20$~GeV) well within the region 
of applicability of perturbative QCD.  The function has essentially no support 
for $b > 0.5$~GeV$^{-1}$.  We expect, therefore, that the non-perturbative 
input at large $b$ will play a negligible role in Higgs boson production 
at LHC energies.  Comparison of Figs.~\ref{fig:bWHLHC}(a) and (b) demonstrates 
that the function is peaked somewhat more narrowly at larger $m_h$.  

We present Fig.~\ref{fig:bWZLHC} to illustrate the 
difference in the resummed $b$ space distributions for Higgs boson and 
$Z$ boson production.  In Figs.~\ref{fig:bWHLHC}(b) and~\ref{fig:bWZLHC}, 
we select the same mass $Q = M_Z$ and collider energy $\sqrt{S}$.  The 
function peaks near $b \sim 0.12$ GeV$^{-1}$ for $Z$ production, at about
twice the value for Higgs boson production.  The peak in the $Q_T$ 
distribution will therefore occur at a smaller value of $Q_T$ for $Z$ 
production than for Higgs boson production.  The 
significantly narrower distribution in $b$ in the Higgs boson case is a 
reflection of the larger gluonic Sudakov factor.  Predictions for the $Q_T$ 
distribution of Higgs boson production are therefore expected to be less 
sensitive to non-perturbative physics than those for $Z$ production, even 
if $m_h = M_Z$.  (We note that at LHC energies, the typical values of $x$ 
in the parton densities are $\sim M_Z/\sqrt{S}$ and $\sim m_h/\sqrt{S}$.  
Even for $m_h = 200$~GeV, $x < 2 \times 10^{-2}$.  Therefore, production at 
the LHC is influenced by the behavior of parton densities at small $x$, 
not at large $x$ where the more rapid fall-off of the gluon density 
might otherwise make the $Q_T$ distribution for Higgs boson production 
decrease more rapidly than that for $Z$ production.)  
  
\subsection{Extrapolation into the region of large-$b$}

The perturbatively resummed function $W(b,Q,x_A,x_B)$ in 
Eq.~(\ref{css-resum}) is reliable only when $b$ is small.  An 
extrapolation into the region of large $b$ is necessary in order to 
complete the Fourier transform in Eq.~(\ref{css-resum}).  
In this section, we summarize different forms used in the literature 
for this extrapolation. The functional form of Qiu and 
Zhang~\cite{Qiu:2000hf} has the desirable property that it separates 
cleanly the perturbative prediction at small $b$ from non-perturbative
physics in the large $b$ region.  

In Ref.~\cite{Collins:1984kg}, a variable $b_*$ is introduced along with  
a non-perturbative function $F^{NP}(b,Q,x_A,x_B)$, 
\begin{equation}
W^{\rm CSS}(b,Q,x_A,x_B) \equiv 
W^{\rm pert}(b_*,Q,x_A,x_B)\,
F^{NP}(b,Q,x_A,x_B)\, ,
\label{css-W-b}
\end{equation}
where $b_*=b/\sqrt{1+(b/b_{\rm max})^2} < b_{\rm max} = 0.5$~GeV$^{-1}$; 
$F^{NP}$ has a Gaussian-like dependence on $b$.  
The assumed non-perturbative distribution 
$F^{NP}$ has the functional form~\cite{Collins:1984kg}
\begin{equation}
F_{ij}^{NP}(b,Q,x_A,x_B) = \exp\left[
-\ln(Q^2 b^2_{\rm max})\, g_1(b) - g_{i/A}(x_A,b) - g_{j/B}(x_B,b) \right] .
\label{css-fnp}
\end{equation}
The $\ln(Q^2)$ dependence is a prediction of the renormalization group 
equations.  The functions
$g_1(b)$, $g_{i/A}(x_A,b)$, and $g_{j/B}(x_B,b)$ are non-perturbative 
and assumed to vanish as $b\rightarrow 0$.  The predictive power of
the CSS formalism relies on the derived $Q^2$ dependence 
and the universality of the functions $F_{ij}^{NP}$, fitted at one 
energy and used elsewhere.  

Davies and Stirling (DS) proposed a simpler form for
$F^{NP}$~\cite{Davies:1984sp} 
\begin{equation}
F^{NP}_{DS}(b,Q,x_A,x_B) 
= \exp\left[-b^2\left(g_1 + g_2 \ln(Qb_{\rm max}/2)\right)\right] \, ,
\label{dws-fnp}
\end{equation}
where $g_1$ and $g_2$ are flavor-independent fitting parameters. 
To incorporate possible $\ln(\tau)$ dependence, $\tau=Q^2/S=x_A x_B$, 
Ladinsky and Yuan (LY) introduced an additional parameter $g_3$ in 
a modified functional form~\cite{Ladinsky:1993zn}
\begin{equation}
F^{NP}_{LY}(b,Q,x_A,x_B) 
= \exp\left[-b^2\left(g_1 + g_2 \ln(Q b_{\rm max}/2)\right) 
- b\,g_1\,g_3\,\ln(100 x_A x_B)\right] \, .
\label{ly-fnp}
\end{equation}
Similar to the DS parameterization, no flavor dependence is 
present.  Landry, Brock, Ladinsky, and Yuan (LBLY)
reported a global fit to the low energy Drell-Yan
data as well as high energy $W$ and $Z$ data using both DS and LY
parameterizations~\cite{Landry:1999an}. 

To preserve the predictive power of perturbative calculations,
it is important that $W^{\rm CSS}(b,Q,x_A,x_B)$ be consistent 
with the perturbatively calculated $W^{\rm pert}(b,Q,x_A,x_B)$ when 
$b<b_{\rm max}$.  However, use of the $b_*$ variable necessarily alters 
the perturbatively calculated $b$-space distribution even within the 
perturbative region~\cite{Qiu:2000hf}.  This alteration can 
be significant depending on the choice of the parametrization of 
$F^{NP}$.  In the last subsection (c.f., Figs.~\ref{fig:bWHLHC} 
and~\ref{fig:bWZLHC} ), we argue 
that the $Q_T$ distributions of $W$, $Z$, and Higgs boson production at 
collider energies should not be sensitive to the large $b$ tail of the 
spectrum.  Any significant dependence on the fitting parameters would cast 
doubt on the predictive power of the $b$-space resummation formalism.    

In Ref.~\cite{Qiu:2000hf}, Qiu and Zhang introduce a new extrapolation 
\begin{equation}
W(b,Q,x_A,x_B) = \left\{
\begin{array}{ll}
 W^{\rm pert}(b,Q,x_A,x_B) &  \mbox{$b\leq b_{\rm max}$} \\
 W^{\rm pert}(b_{\rm max},Q,x_A,x_B)\,
 F^{NP}(b,Q;b_{\rm max})
                           &  \mbox{$b > b_{\rm max}$}.
\end{array} \right.
\label{qz-W-b}
\end{equation}
This expression manifestly preserves the QCD resummed $b$-space distribution 
in the perturbative region at small $b$.  The authors also motivate the adoption 
of a new non-perturbative function in the region of large $b$: 
\begin{eqnarray}
F^{NP}
&=&
\exp\bigg\{
 -\ln(\frac{Q^2 b_{\rm max}^2}{c^2}) \left[
   g_1 \left( (b^2)^\alpha - (b_{\rm max}^2)^\alpha\right)
  +g_2 \left( b^2 - b_{\rm max}^2\right) \right]
\nonumber \\
&\ & \hskip 0.4in
 -\bar{g}_2 \left( b^2 - b_{\rm max}^2\right)
  \bigg\} .
\label{qz-fnp}
\end{eqnarray}
The term proportional to $(b^2)^\alpha$, with $\alpha<1/2$, corresponds to 
a direct extrapolation of the resummed $W^{\rm pert}(b,Q,x_A,x_B)$.
The parameters, $g_1$ and $\alpha$ are fixed in relation to $W^{\rm pert}$ by 
the requirement that the first and second derivatives of
$W(b,Q,x_A,x_B)$ be continuous at $b=b_{\rm max}$. Dependence on $x_A$ and 
$x_B$ is included in the parameters $g_1$ and $\alpha$.  
The terms proportional to $b^2$ correspond to inverse-power corrections to 
the evolution equation. Power corrections from soft gluon showers are the 
origin of the $g_2$ term, and non-vanishing intrinsic transverse momenta  
of the incident partons lead to the $\bar{g}_2$ term. The values of 
$g_2$ and $\bar{g}_2$ are unknown and must be obtained from fits to data.
\begin{figure}[ht]
\centerline{\includegraphics[width=9.0cm]{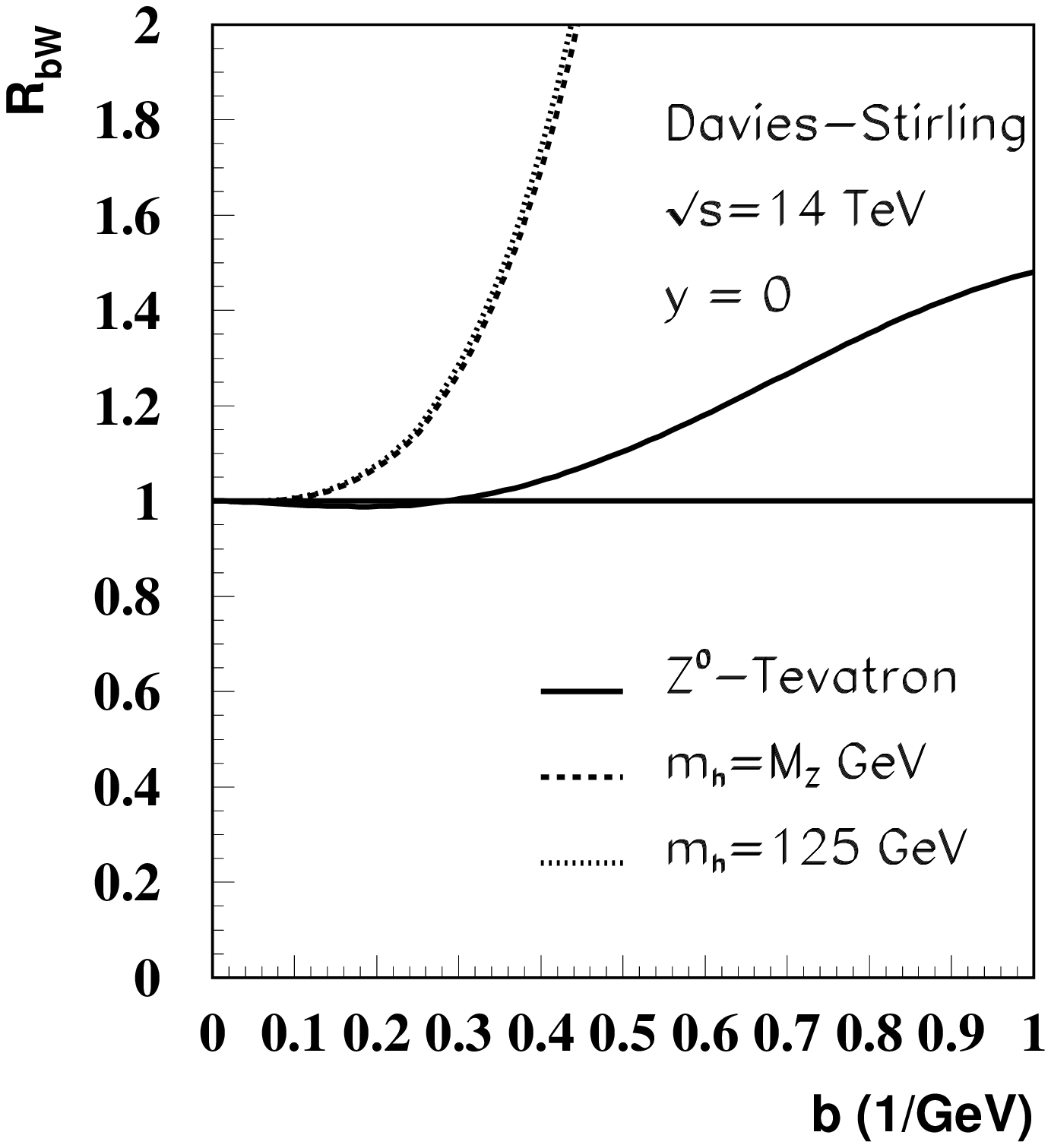}
\includegraphics[width=9.0cm]{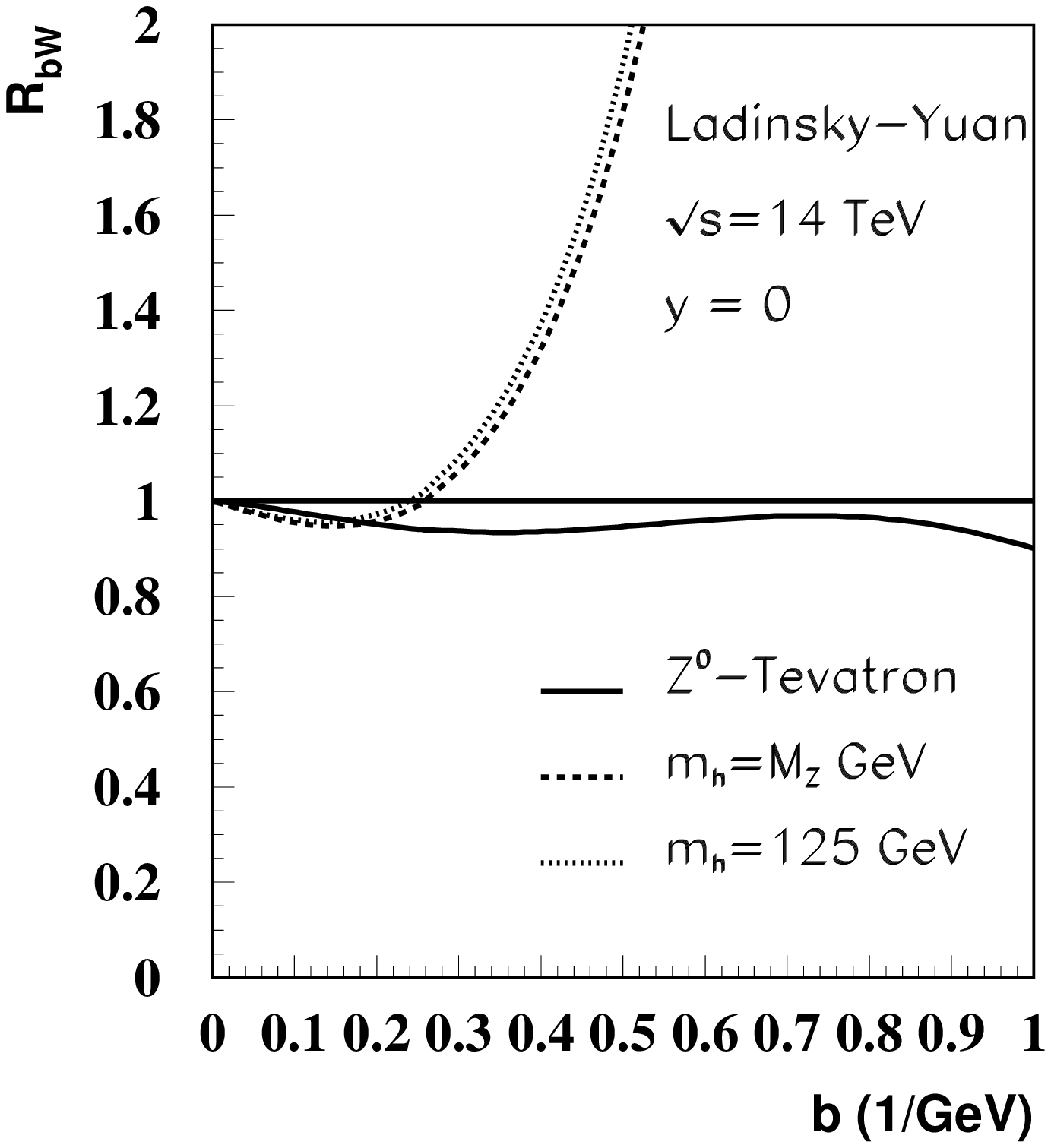}} 
\caption[]{\it Ratios of the functions $bW(b,Q)$.  Comparisons 
  are shown of (a) the DS parametrization and (b) the LY choice with 
  respect to the function we use in this paper.  The solid 
  lines show the comparisons for $Z$ production at the energy of the 
  Fermilab Tevatron, $\sqrt S = 1.8$~TeV.   
  The dashed and dotted lines show the ratios for Higgs boson production 
  at the LHC.} 
\label{fig:bWratios}
\end{figure}

In Fig.~\ref{fig:bWratios} we present ratios of the functions $bW(b,Q)$  
in use by different groups.  In Fig.~\ref{fig:bWratios}(a), the numerator 
is the two-parameter form of Davies and Stirling, 
whereas in Fig.~\ref{fig:bWratios}(b) it is the three-parameter functional 
form adopted by Ladinsky and Yuan, 
Eqs.~(\ref{css-W-b}) and (\ref{ly-fnp}).  However, in both cases we use 
the updated parameters of Landry {\em et al} and the CTEQ5M parton 
densities.  The denominator in the ratios is the Qiu and Zhang function we 
use in this 
paper, Eqs.~(\ref{qz-W-b}) and (\ref{qz-fnp}), with $b_{\rm max} = 0.5$, 
$g_2 =0$, and $\bar{g}_2 =0$.  For $Z$ production at the energy of the 
Tevatron, the ratios do not deviate from 1 by more than 20\% as long 
as $b < 0.5$, as might be expected because all groups provide acceptable 
fits to data.  However, very significant deviations from 1 are evident at 
the energy of the LHC, notably in the region $b < 0.5$ where purely 
perturbative physics should be dominant.  The deviations reach 100\% for 
$b \sim 0.4$.  This comparison shows explicitly the influence within the 
purely perturbative region of the use of the variable $b_*$ and the 
nonperturbative function $F^{NP}$ in Eq.~(\ref{css-W-b}).

Figure~\ref{fig:bWratios} shows that the deviations from the 
perturbatively calculated $W^{\rm pert}(b,Q)$ depend strongly on 
the collision energy $\sqrt{S}$.  This energy dependence is an 
artifact of the use of the variable $b_*$.  Since $b_* < b$ for
all $b\neq 0$, the shifted $b$-space distribution $W^{\rm pert}(b_*,Q) 
< W^{\rm pert}(b,Q)$ for values of $b$ smaller than the location of 
saddle point, and $W^{\rm pert}(b_*,Q) > W^{\rm pert}(b,Q)$ for $b$ 
larger than the saddle point.  The magnitude of the difference between the
$W^{\rm pert}(b_*,Q)$ and $W^{\rm pert}(b,Q)$ depends on the shape of
$W^{\rm pert}(b,Q)$.   When $\sqrt{S}$ increases, $W^{\rm pert}(b,Q)$ 
becomes narrower, and its much steeper behavior leads to a larger 
deviation of $W^{\rm pert}(b_*,Q)$ from $W^{\rm pert}(b,Q)$.  
The use of either the DS or the LY parameterizations in 
$W^{\rm CSS}(b,Q)$, Eq.~(\ref{css-W-b}), leads to a reduction 
in the region $b < 0.2$ 
and strong enhancement in the region $b > 0.2$.  Once the Fourier 
transform is made to $Q_T$ space, the net effects of these two 
differences are a slight shift to smaller $Q_T$ in the location of 
the predicted maximum of the distribution $d\sigma/dydQ_T$ and 
reduction in the magnitude of the peak value.

\section{Expressions for the fixed-order contributions in 
perturbative QCD}

Working at fixed-order in QCD perturbation 
theory~\cite{Ellis:1987xu,deFlorian:1999zd,Ravindran:2002dc,Glosser:2002gm},
one can express the differential cross section for Higgs boson production in 
the form of a convolution  
\begin{eqnarray}
\frac{d\sigma_{AB\rightarrow h X}^{\rm (pert)}}{dQ^2\, dy\, dQ_T^2}
&=&
\sigma_{gg\rightarrow h X}^{(0)}
\sum_{ab}
\int_{x_A}^1  \frac{d\xi_A}{\xi_A}\,
\int_{x_B}^1  \frac{d\xi_B}{\xi_B}\,
\phi_{a/A}(\xi_A,\mu)\,\phi_{b/B}(\xi_B,\mu)
\nonumber\\
&\ & \hskip 0.4in \times
H_{ab\rightarrow h X}(Q_T,Q,\frac{x_A}{\xi_A},\frac{x_B}{\xi_B},\alpha_s(\mu),\mu) .
\label{fo1}
\end{eqnarray}
In this expression, $\sigma_{gg\rightarrow h X}^{(0)}$ is the cross 
section given in Eq.~(\ref{gg-H-lo}). The functions $\phi_{a/A}(x_A,\mu)$ 
and $\phi_{b/B}(x_B,\mu)$ are 
probabilities that incident partons $a$ and $b$ from 
hadrons $A$ and $B$ carry fractional light-cone momenta $x_A$ and 
$x_B$, respectively, at factorization scale $\mu$.  The parton level  
hard-scattering function 
$H_{ab\rightarrow h X}(Q_T,Q,x_A/\xi_A,x_B/\xi_B,\alpha_s(\mu),\mu)$   
is expanded in a power-series in the strong coupling strength 
$\alpha_s$
\begin{equation}
H_{ab\rightarrow h X}=\sum_{n=0}
H_{ab\rightarrow h X}^{(n)} \left(\frac{\alpha_s(\mu)}{\pi}\right)^n .  
\end{equation}
\noindent
The lowest order contribution ($n = 0$), $gg \rightarrow h$ comes from 
the top-quark loop, and, in the absence of partonic intrinsic transverse 
momentum, it yields a distribution that is a $\delta$-function in 
transverse momentum $Q_T$  
\begin{equation}
H_{gg}^{(0)} = \delta\left(1-\frac{x_A}{\xi_A}\right)\,
               \delta\left(1-\frac{x_B}{\xi_B}\right)\,
               \delta^2(Q_T) .
\label{ggzero}
\end{equation}
\noindent
At first-order in $\alpha_s$, there are virtual loop-corrections to 
the subprocess $gg \rightarrow h$ along with contributions in which 
two partons interact to produce a parton plus the Higgs boson.  
Labeling these 2-to-2 contributions by the tree-subprocesses, we 
write the functions 
$H_{ab\rightarrow h X}^{(1)}$ in terms of partonic Mandelstam variables 
$\hat{s}$, $\hat{t}$, and $\hat{u}$ as~\cite{Ellis:1987xu} 

\noindent
{\em gluon-gluon} $gg \rightarrow hg$:

\begin{equation}
H_{gg}^{(1)} = \frac{1}{2\pi}\,C_A
\left[\frac{\hat{s}^4+\hat{t}^4+\hat{u}^4+Q^8}
           {\hat{s}\,\hat{t}\,\hat{u}\,Q^2} 
\right]\,
\delta\left( \hat{s}+\hat{t}+\hat{u}-Q^2 \right) ,
\label{ggone}
\end{equation}

\noindent
{\em gluon-quark(antiquark)} $g q \rightarrow hq$ 
($g \bar{q} \rightarrow h \bar{q}$):

\begin{equation}
H_{gq}^{(1)} = \frac{1}{2\pi}\,C_F
\left[\frac{\hat{s}^2+\hat{u}^2}
           {-\hat{t}\,Q^2} 
\right]\,
\delta\left( \hat{s}+\hat{t}+\hat{u}-Q^2 \right) ,
\end{equation}

\noindent
{\em quark(antiquark)-gluon} $q g \rightarrow hq$ 
($\bar{q} g \rightarrow h \bar{q})$: 

\begin{equation}
H_{qg}^{(1)} = \frac{1}{2\pi}\,C_F
\left[\frac{\hat{s}^2+\hat{t}^2}
           {-\hat{u}\,Q^2} 
\right]\,
\delta\left( \hat{s}+\hat{t}+\hat{u}-Q^2 \right) ,
\end{equation}
and the s-channel {\em quark-antiquark} subprocess 
$q \bar{q} \rightarrow hg$:

\begin{equation}
H_{q\bar{q}}^{(1)} = \frac{1}{2\pi}
\left(\frac{C_F}{C_A}\left(N^2-1\right)\right) 
\left[\frac{\hat{t}^2+\hat{u}^2}
           {\hat{s}\,Q^2} 
\right]\,
\delta\left( \hat{s}+\hat{t}+\hat{u}-Q^2 \right) .
\label{qqbarone}
\end{equation}

The first-order contributions from $gg$, $qg$, and $gq$ scattering are 
all singular as $Q_T \rightarrow 0$.  Following convention, we use 
the word ``asymptotic'' to designate the terms that are at least as singular 
as $Q_T^{-2}$. This 
asymptotic contribution may be written in analogy to Eq.~(\ref{fo1}) as 
the convolution 
\begin{eqnarray}
\frac{d\sigma_{AB\rightarrow h X}^{\rm (asym)}}{dQ^2\, dy\, dQ_T^2}
&=&
\sigma_{gg\rightarrow h X}^{(0)}
\sum_{ab}
\int_{x_A}^1 \frac{d\xi_A}{\xi_A}\,
\int_{x_B}^1 \frac{d\xi_B}{\xi_B}\,
\phi_{a/A}(\xi_A,\mu)\,\phi_{b/B}(\xi_B,\mu)
\nonumber\\
&\ & \hskip 0.4in \times
w_{ab\rightarrow h X}(Q_T,Q,\frac{x_A}{\xi_A},\frac{x_B}{\xi_B},\alpha_s(\mu),\mu) .
\end{eqnarray}
Expanded in a power series in $\alpha_s$, the asymptotic partonic hard part 
$w_{ab\rightarrow h X}$ becomes 
\begin{equation}
w_{ab\rightarrow h X}=\sum_{n=1}
w_{ab\rightarrow h X}^{(n)} \left(\frac{\alpha_s(\mu)}{\pi}\right)^n .
\end{equation}
The lowest order asymptotic contributions ($n = 0$) vanish for $Q_T > 0$, 
\begin{equation}
w_{ab}^{(0)} = 0 .
\label{ggaszero}
\end{equation}

\noindent
Categorizing the first-order contributions as before, we write the 
asymptotic forms as  

\noindent
{\em gluon-gluon:}

\begin{eqnarray}
w_{gg}^{(1)} 
&=& 
\frac{1}{2\pi}\,\frac{1}{Q_T^2}
\bigg\{2\delta(1-z_A)\delta(1-z_B)
     \left[A_g^{(1)}\log(\frac{Q^2}{Q_T^2})+B_g^{(1)}\right]
\nonumber \\
&\ & \hskip 0.5in
      +\delta(1-z_A)\, P_{g\rightarrow g}(z_B)
      +\delta(1-z_B)\, P_{g\rightarrow g}(z_A) 
\bigg\} ,
\label{ggasone}
\end{eqnarray}
where 
\begin{equation}
z_A = \frac{x_A}{\xi_A},
\quad\mbox{and}\quad
z_B = \frac{x_B}{\xi_B} .
\end{equation}

\noindent
The forms are simpler for the other subprocesses. 

\noindent  
{\em gluon-quark(antiquark):}

\begin{equation}
w_{gq}^{(1)} = \frac{1}{2\pi}\,\frac{1}{Q_T^2}
\left\{\delta(1-z_A)\, P_{q\rightarrow g}(z_B)
\right\} ,
\end{equation}

\noindent
{\em quark(antiquark)-gluon:}

\begin{equation}
w_{qg}^{(1)} = \frac{1}{2\pi}\,\frac{1}{Q_T^2}
\left\{\delta(1-z_B)\, P_{q\rightarrow g}(z_A)
\right\} ,
\end{equation}
and 
\noindent
{\em quark-antiquark:}

\begin{equation}
w_{q\bar{q}}^{(1)} = 0 .  
\label{qqbarasone}
\end{equation}

As discussed in Sec.~II, resummation treats only the parts of the 
fixed-order QCD expressions that are at least as singular as 
$Q_T^{-2}$ in the limit $Q_T \rightarrow 0$.  The remainder is   
\begin{eqnarray}
\frac{d\sigma_{AB\rightarrow h X}^{\rm (Y)}}{dQ^2\, dy\, dQ_T^2}
&=&
\sigma_{gg\rightarrow h X}^{(0)}
\sum_{ab}
\int_{x_A}^1  \frac{d\xi_A}{\xi_A}\,
\int_{x_B}^1  \frac{d\xi_B}{\xi_B}\,
\phi_{a/A}(\xi_A,\mu)\,\phi_{b/B}(\xi_B,\mu)
\nonumber\\
&\ & \hskip 0.4in \times
R_{ab\rightarrow h X}(Q_T,Q,\frac{x_A}{\xi_A},\frac{x_B}{\xi_B},\alpha_s(\mu),\mu) .
\label{yterm}
\end{eqnarray}

\noindent
As a power-series expansion in $\alpha_s$, the partonic hard part for 
the remainder term is 

\begin{equation}
R_{ab\rightarrow h X}=\sum_{n=0}
R_{ab\rightarrow h X}^{(n)} \left(\frac{\alpha_s(\mu)}{\pi}\right)^n , 
\end{equation}
where
\begin{equation}
R_{ab\rightarrow h X}^{(n)}
=
H_{ab\rightarrow h X}^{(n)}
-
w_{ab\rightarrow h X}^{(n)} .  
\end{equation}
In our numerical work presented in Sec.~V, we use order $\alpha_s$ 
expressions for $H_{ab\rightarrow h X}^{(n)}$.  In subsequent investigations, 
one might incorporate the NLO results~\cite{Ravindran:2002dc,Glosser:2002gm}.  
   
\section{Predictions}
In this section we present our predictions for the transverse momentum 
distribution of Higgs boson production in proton-proton collisions 
at the LHC energy $\sqrt{S} = 14$~TeV.  We provide results at both 
fixed rapidity $y = 0$ and for the cross section integrated over the 
rapidity range $|y| \le 2.4$. To illustrate interesting differences, 
we also show our results for $Z$ boson production at the same 
energy~\cite{Zhang:2002yz}.  
\begin{figure}[ht]
\centerline{\includegraphics[width=9.0cm]{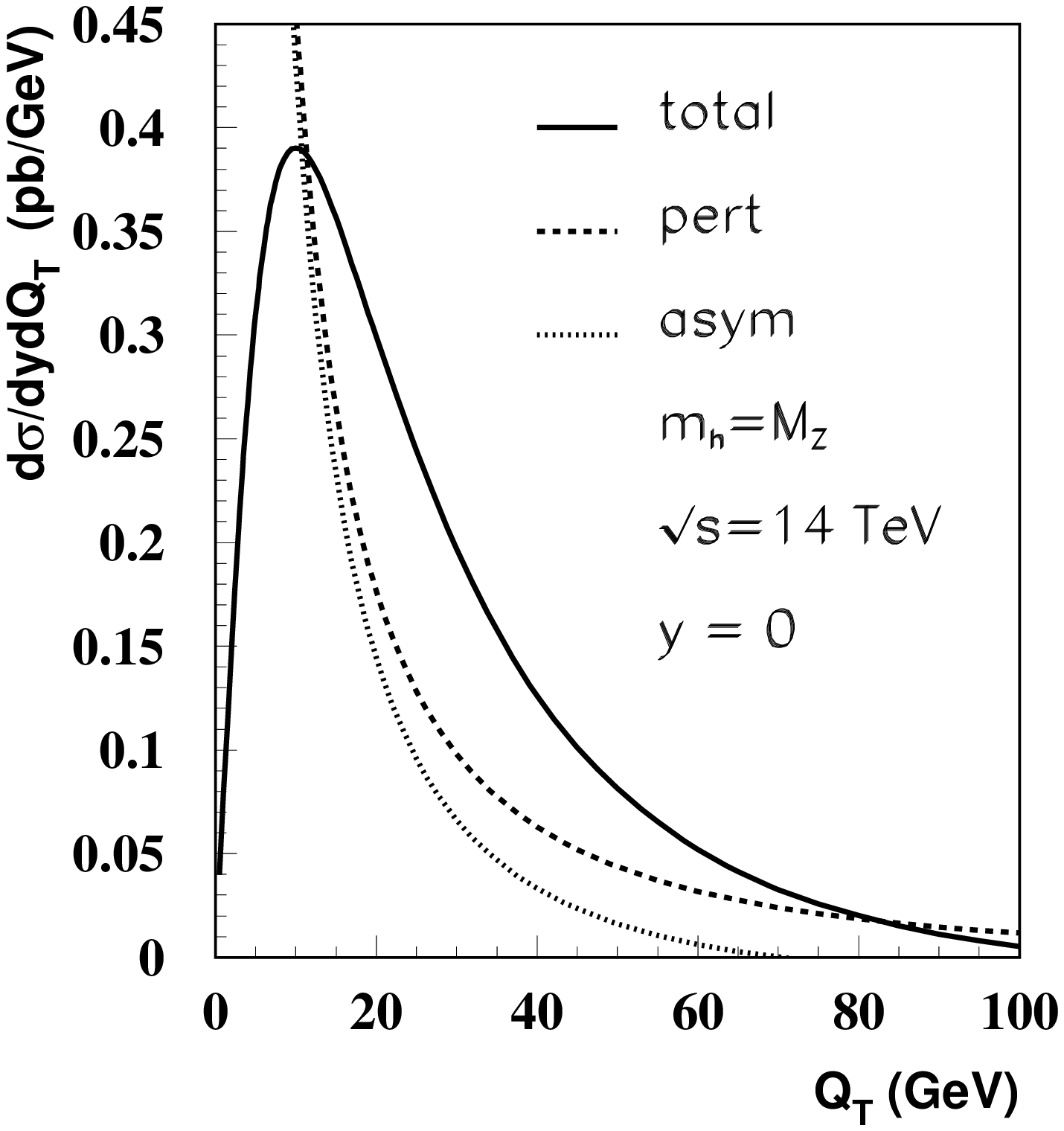}
\includegraphics[width=9.0cm]{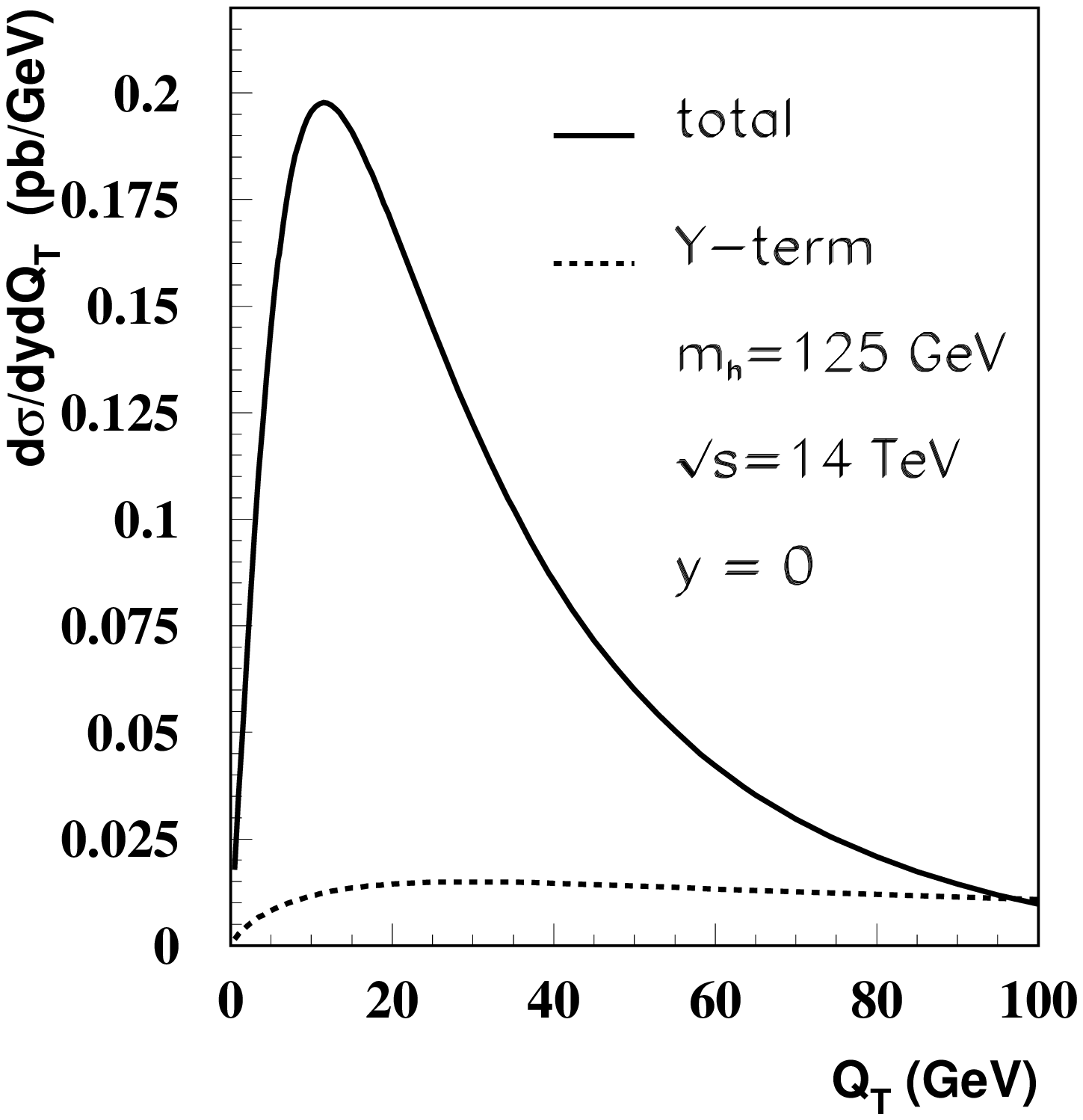}} 
\centerline{\includegraphics[width=9.0cm]{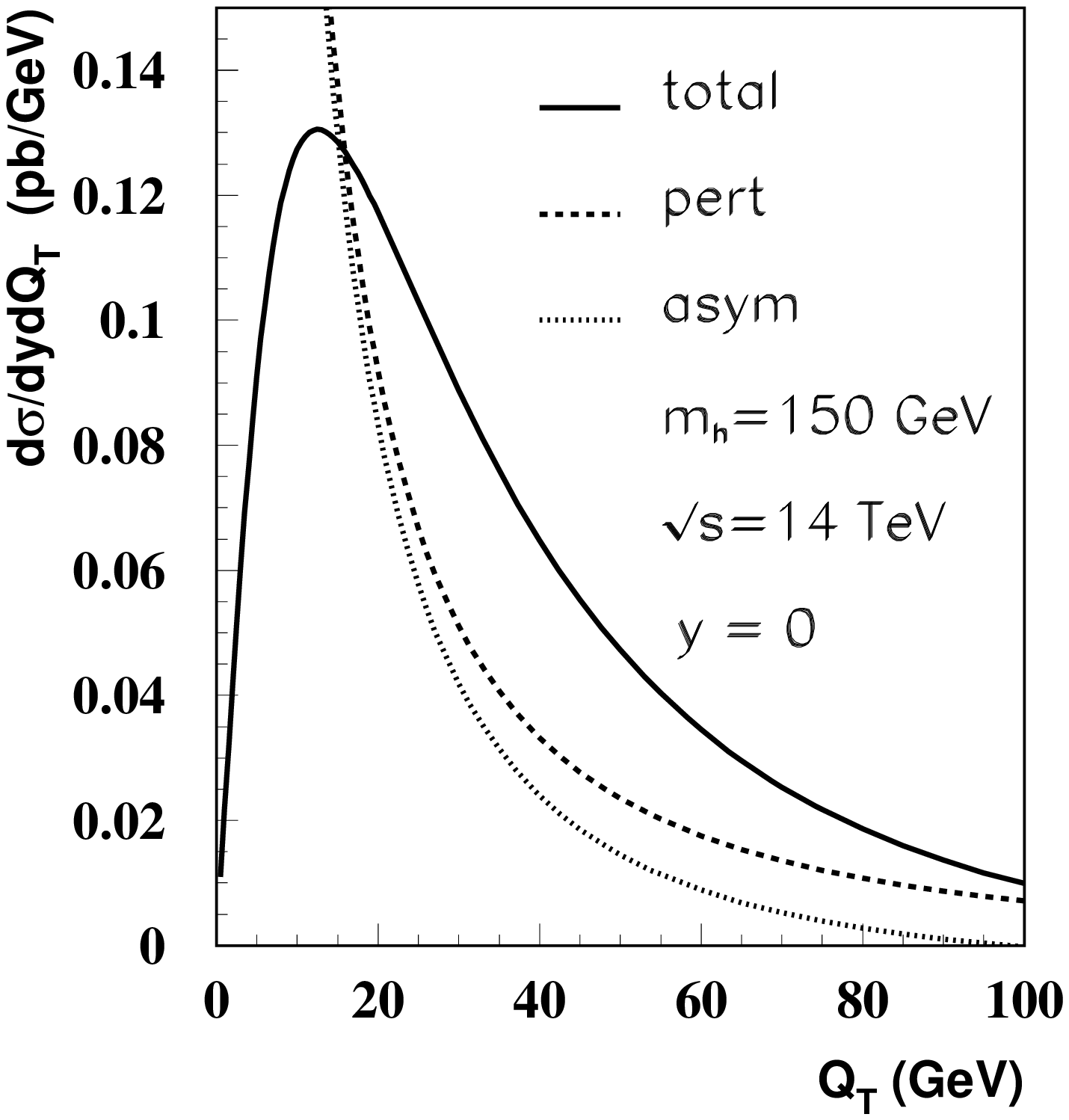}
\includegraphics[width=9.0cm]{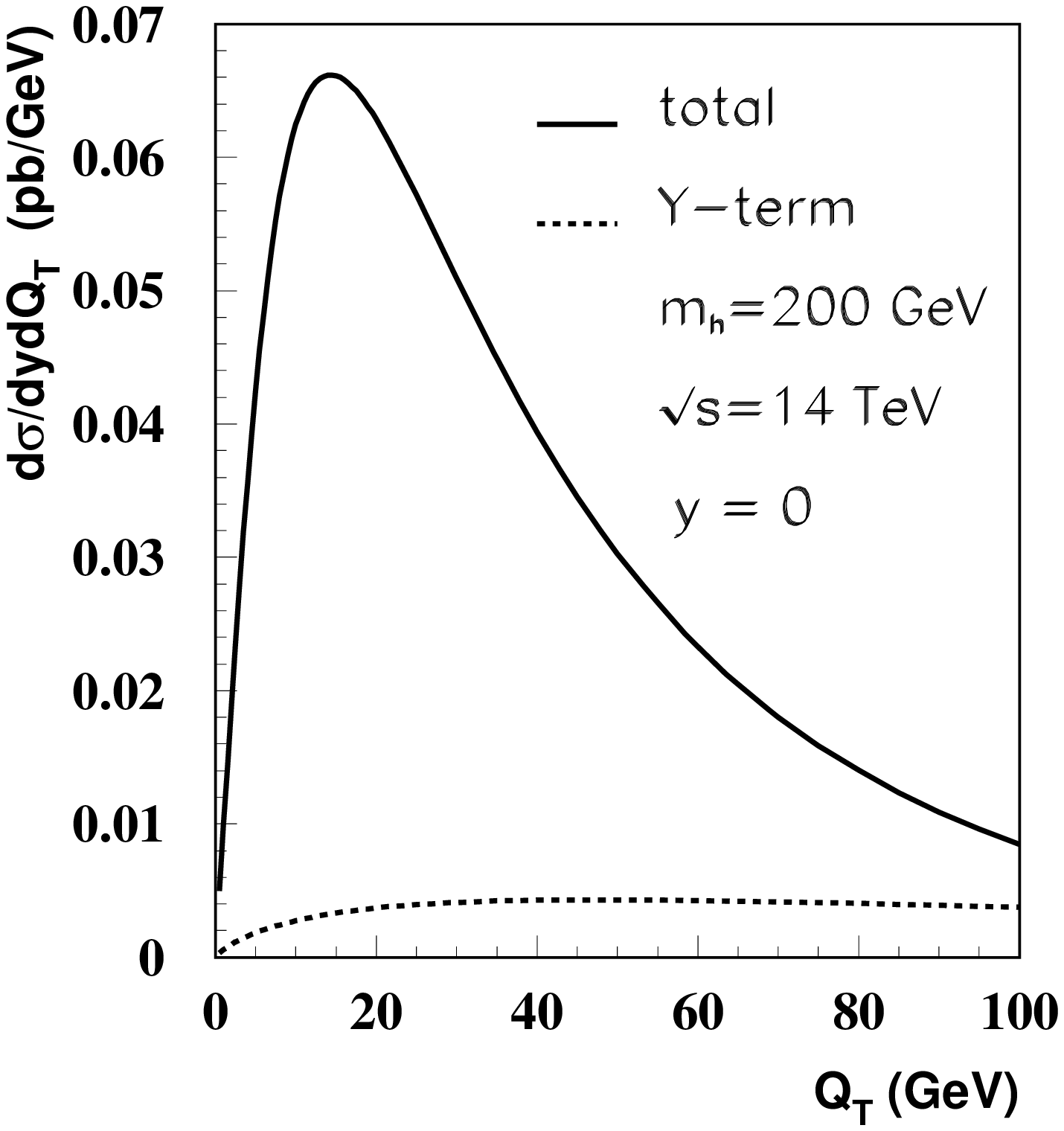}} 
\caption[]{\it Differential cross section $d\sigma/dy dQ_T$ for Higgs boson 
  production at $\sqrt{S} = 14$~TeV and fixed rapidity $y=0$ with 
  (a) $Q= M_Z$, (b) $Q=m_h=125$ GeV, (c) $Q=m_h= 150$ GeV, and (d) 
  $Q=m_h=200$ GeV, as a function 
  of transverse momentum $Q_T$. The total prediction is shown as a 
  solid line.  In (a) and (c), the fixed-order perturbative calculation 
  is shown as a dashed line and its asymptotic limit as a dotted line.  In 
  (b) and (d), the contribution from the remainder $Y$ term is shown as a 
  dashed line.}
\label{fig:QTHiggs0}
\end{figure}
Our predictions are based on 
Eqs.~(\ref{css-gen}),~(\ref{css-resum}),~(\ref{css-pert}), and~(\ref{yterm}). 
We employ expressions for the 
parton-level hard-scattering functions valid through first-order 
in $\alpha_s$, Eqs.~(\ref{ggzero}) - (\ref{qqbarone}), with the 
asymptotic forms Eqs.~(\ref{ggaszero}) - (\ref{qqbarasone}).  
For $A_g$ and $B_g$ in the Sudakov factor, Eq.~(\ref{css-S}), 
we take the expansion valid through second order ($n = 2$), 
Eq.~(\ref{css-AB}). For the coefficient functions in the modified 
parton densities, we use the expansion through $n = 1$, 
Eq.~(\ref{css-coef}).  (Our results are therefore consistently at 
next-to-leading-logarithm (NLL) accuracy).  We use a next-to-leading 
order form for $\alpha_s(\mu)$ and next-to-leading order normal parton  
densities $\phi(x,\mu)$~\cite{Lai:1999wy}.  In the fixed-order 
perturbative expressions that enter the ``$Y$'' function, we select a 
fixed renormalization/factorization scale 
$\mu = \kappa \sqrt{m_h^2 + Q_T^2}$, with $\kappa = 0.5$.  For the 
resummed term, we take as our central value $\mu = c/b$, with 
$c=2e^{-\gamma_E}$.  To show sensitivity to the choice of $\mu$, 
we also compute cross sections with the choices 
$\mu = 0.5 m_h$ and $\mu = 2 m_h$.  

The extrapolation into the non-perturbative region of large $b$ is 
accomplished with Eqs.~(\ref{qz-W-b}) and (\ref{qz-fnp}).  
As remarked before, the parameters $g_1$ and $\alpha$ in Eq.~(\ref{qz-fnp}) 
are fixed in relation to $W^{\rm pert}$ by the requirement that the first 
and second derivatives of $W(b,Q,x_A,x_B)$ be continuous at 
$b=b_{\rm max}=0.5$~GeV$^{-1}$.  Since, as shown in Sec.~III~A, there is 
little sensitivity to the region of large $b$, we adopt for most of 
our results the values $g_2 = 0$ and $\bar{g}_2 = 0$ in Eq.~(\ref{qz-fnp}).  
However, to demonstrate the insensitivity numerically, we also present 
results for other choices of $b_{\rm max}$, $g_2$, and $\bar{g}_2$.  

In the limit $Q_T \rightarrow 0$, sub-dominant singularities may produce a 
mismatch~\cite{Collins:1984kg} between the complete fixed-order expressions 
in Eqs.~(\ref{ggone})~-~(\ref{qqbarone})  
and their asymptotic forms in Eqs.~(\ref{ggasone})~-~(\ref{qqbarasone}).
Sub-leading divergences could drive the difference toward a negative 
infinity 
as $Q_T \rightarrow 0$.  The ``$Y$'' term is not large in the region of 
small $Q_T$, but the singularity in the limit $Q_T \rightarrow 0$ can 
pull the sum in Eq.~(\ref{css-gen})
to negative values in the region of very small $Q_T$.  There are various 
possible remedies for this difficulty, including introduction of an 
off-set in $Q_T$ in the ``$Y$'' function.  In our numerical work, we 
use a parameter $Q_T^{\rm min}$, replacing $Q_T$ in the ``$Y$'' function 
only by $\sqrt{Q^2_T + (Q_T^{\rm min})^2}$.  In keeping with CSS, we select 
$Q_T^{\rm min} = 0.3$~GeV as our central value and show the sensitivity 
of results to choices ranging from $Q_T^{\rm min} =0$ to 
$Q_T^{\rm min} = 0.5$~GeV.  

The Bessel function in the integrand of Eq.~(\ref{css-resum}) is an 
oscillatory function, and a high degree of numerical accuracy is necessary 
to ensure accurate cancellations.  The number of oscillations 
depends strongly on the value of $Q_T$. When $b \in (0, 2)$~GeV$^{-1}$, 
the number of times $J_0(Q_T b)$ crosses zero is 0, 6, 63, and 127
for $Q_T=1$, 10, 100, and 200~GeV, respectively.  Following Qiu and Zhang, 
we use an integral form for the Bessel function 
\begin{equation}
J_0(z) = \frac{1}{\pi}\, 
\int_0^\pi \cos\left(z\sin(\theta)\right) d\theta\, .
\label{bessel}
\end{equation}
Overall numerical accuracy is controlled by improving the accuracy of the 
integration in Eq.~(\ref{bessel}).

\subsection{$Q_T$ distributions for Higgs boson and $Z$ production}
\label{sec5a}

\begin{figure}[ht]
\centerline{\includegraphics[width=9.0cm]{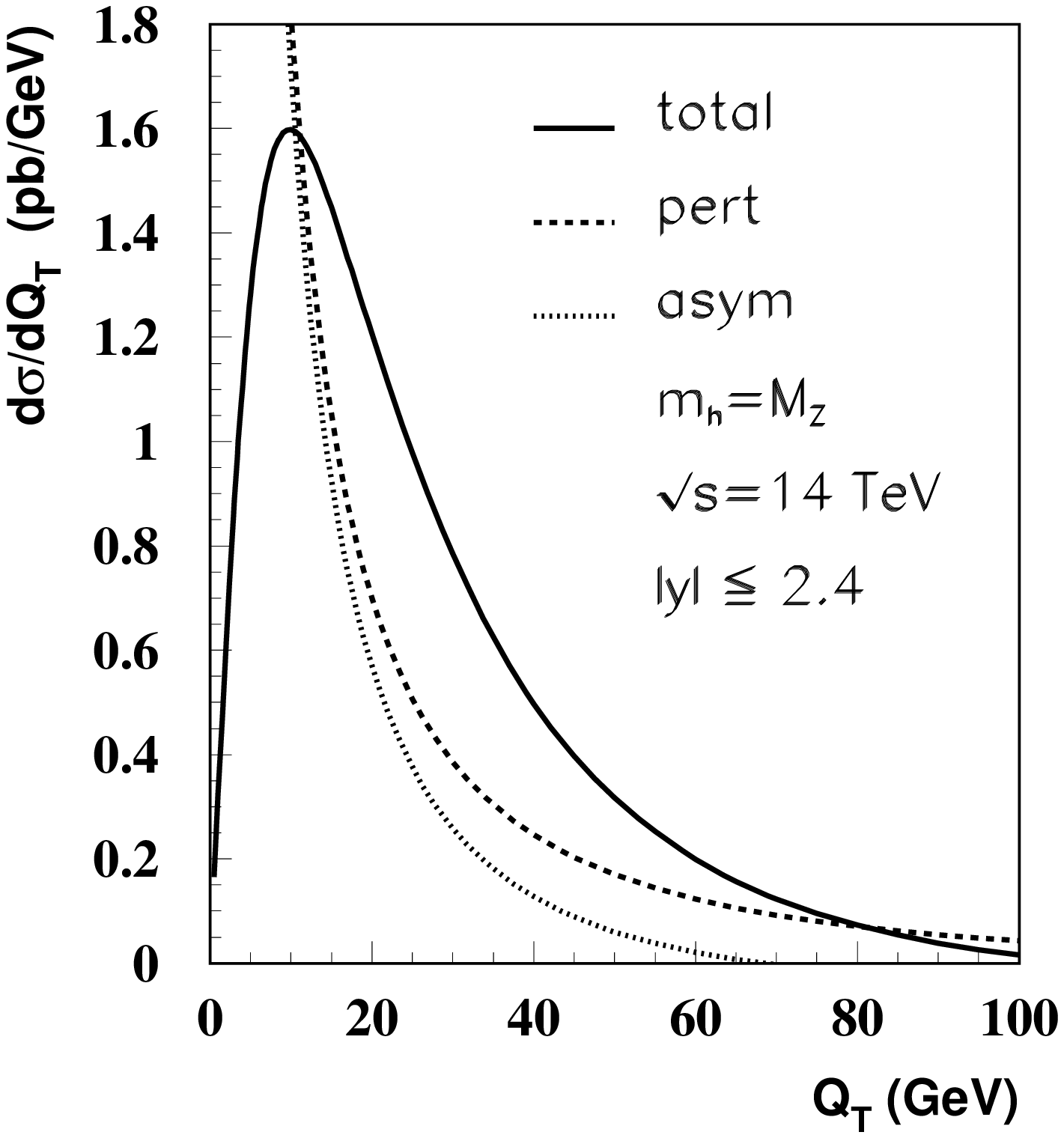}
\includegraphics[width=9.0cm]{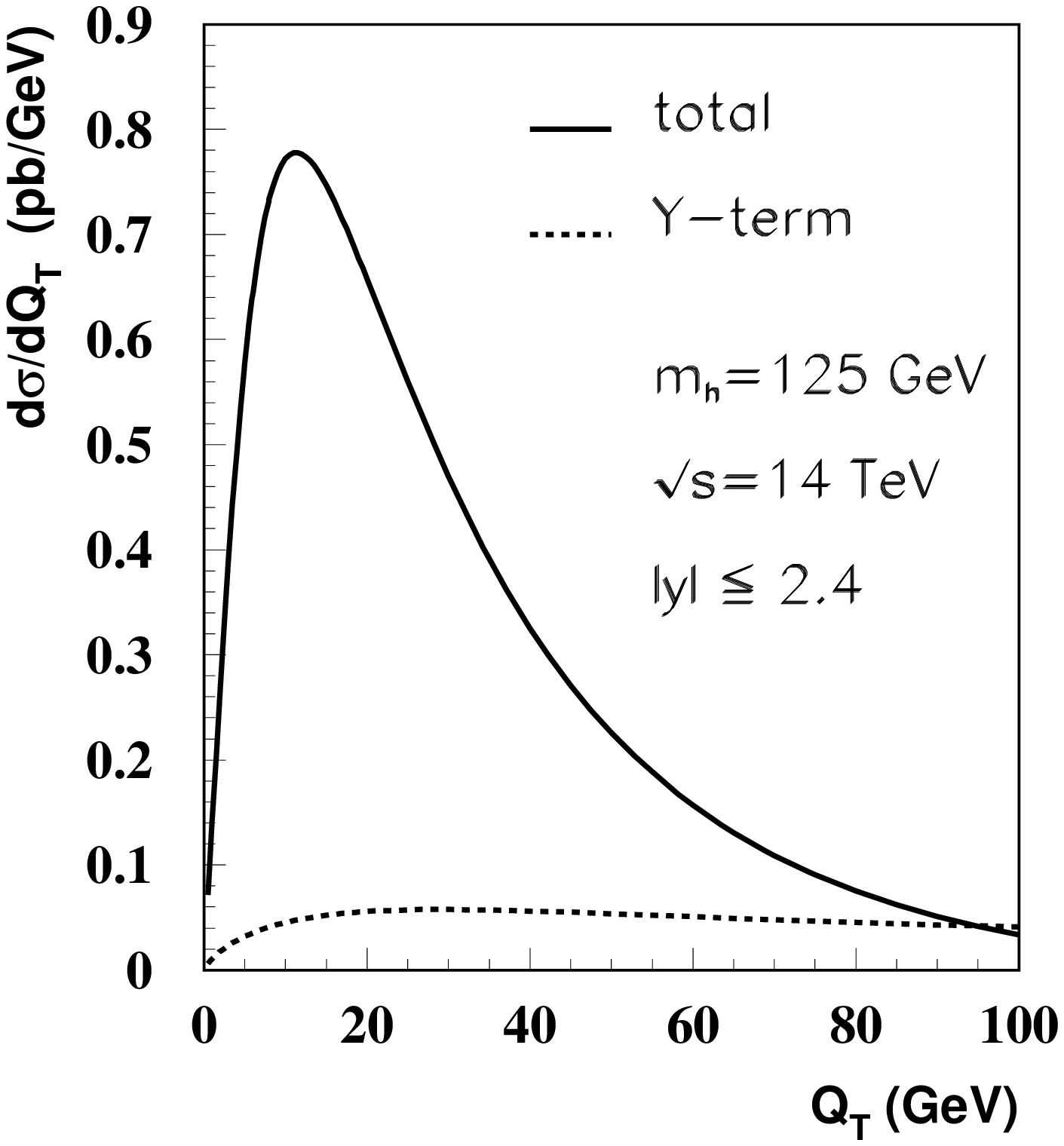}} 
\centerline{\includegraphics[width=9.0cm]{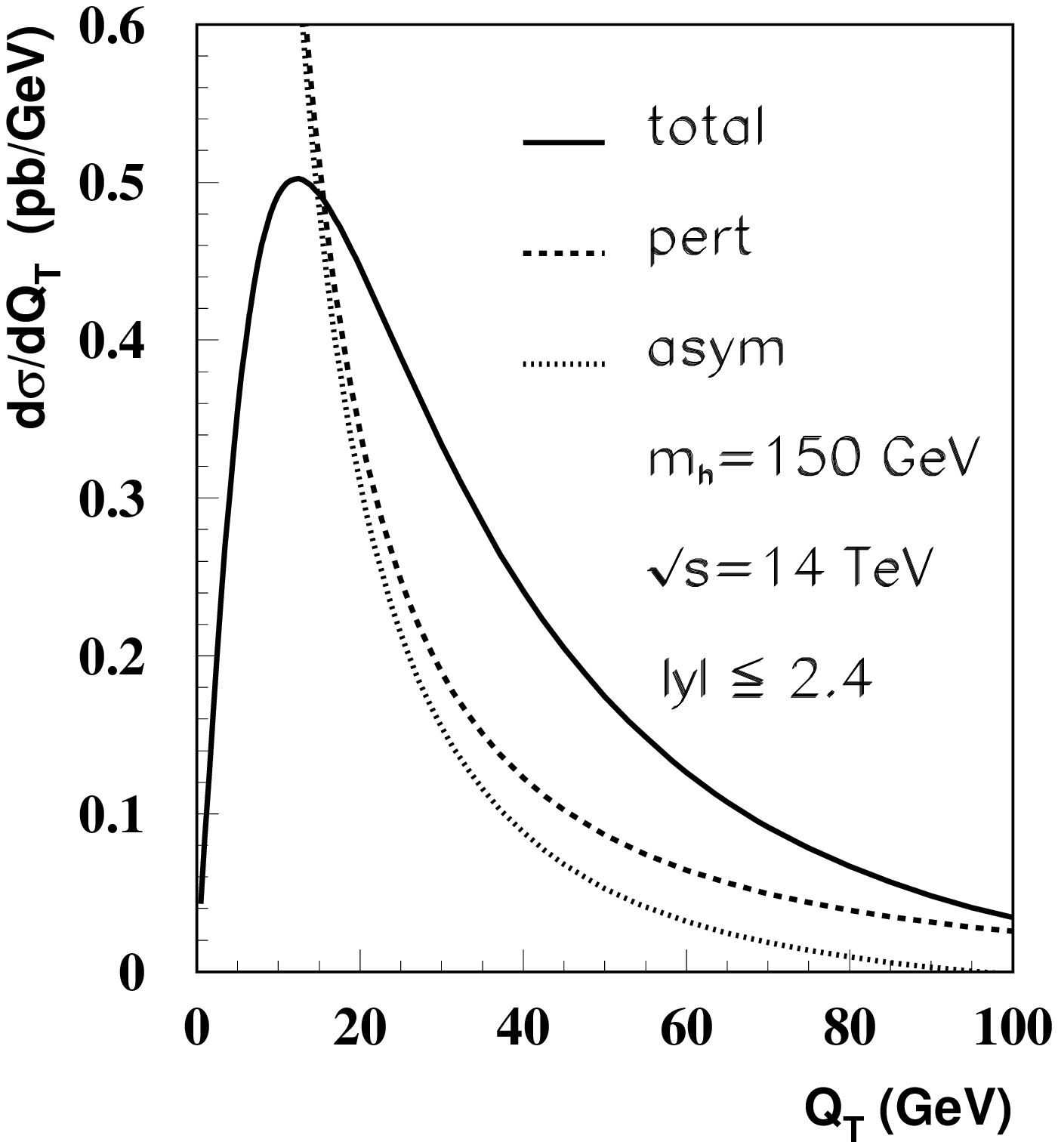}
\includegraphics[width=9.0cm]{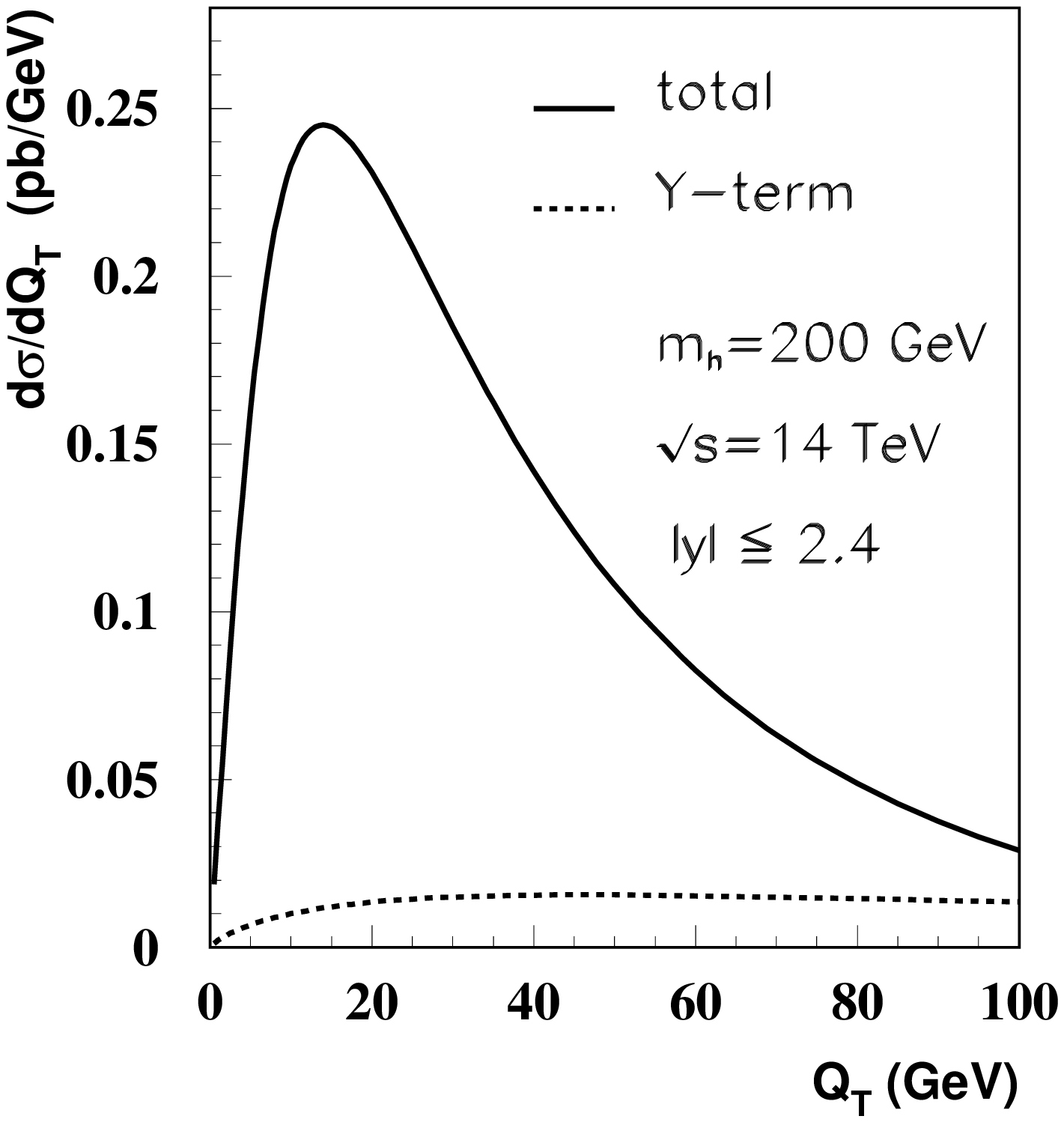}} 
\caption[]{\it Differential cross section $d\sigma/dQ_T$ for Higgs boson 
  production at $\sqrt{S} = 14$~TeV with rapidity integrated over the 
  interval $|y| < 2.4$ with 
  (a) $Q= M_Z$, (b) $Q=m_h=125$ GeV, (c) $Q=m_h= 150$ GeV, and (d) 
  $Q=m_h=200$ GeV, as a function 
  of transverse momentum $Q_T$. The total prediction is shown as a 
  solid line.  In (a) and (c), the fixed-order perturbative calculation 
  is shown as a dashed line and its asymptotic limit as a dotted line.  In 
  (b) and (d), the contribution from the remainder $Y$ term is shown as a 
  dashed line.}
\label{fig:QTHiggs24}
\end{figure}

We display our predicted $Q_T$ distributions for Higgs boson production 
at the LHC in Figs.~\ref{fig:QTHiggs0} and \ref{fig:QTHiggs24}, respectively. 
In Fig.~\ref{fig:QTHiggs0}, we show $d\sigma/dy dQ_T$ as a function of $Q_T$ at
fixed rapidity $y = 0$, whereas in Fig.~\ref{fig:QTHiggs24}, we show the $Q_T$ 
distributions for rapidity integrated over the interval $|y| < 2.4$.  (The 
choice $|y| < 2.4$ is motivated by the expected acceptance of the LHC 
detectors for photons~\cite{atlas:1999fr}.)  We 
present results for four choices of mass of the Higgs boson, values that 
span the range of present interest in the SM, $m_h = M_Z = 91.187$~GeV, 
$m_h = 125$ GeV, $m_h = 150$~GeV (where the $W W^*$ decay channel is 
dominant), and $m_h = 200$~GeV (above both $WW$ and $ZZ$ decay thresholds).  
In all cases, the solid lines represent the total prediction, 
Eq.~(\ref{css-gen}). In Figs.~\ref{fig:QTHiggs0}(a) and (c), as well as 
in Figs.~\ref{fig:QTHiggs24}(a) and (c), we also show the fixed-order 
purely perturbative result and its $Q_T \rightarrow 0$ asymptotic form.  
In Figs.~\ref{fig:QTHiggs0}(b) and (d), as well as in 
Figs.~\ref{fig:QTHiggs24}(b) and (d), we show instead the contribution 
from the fixed-order remainder ``$Y$'' term.  

In Figs.~\ref{fig:QTHiggs0}(a) and (c), we observe the divergence as 
$Q_T \rightarrow 0$ of the fixed-order purely perturbative result and 
the expected numerical equality of the purely perturbative result and 
its $Q_T \rightarrow 0$ asymptotic form at small $Q_T$.  The asymptotic 
form falls away more rapidly with $Q_T$.  Figures~\ref{fig:QTHiggs0}(b) 
and (d) demonstrate that the total prediction, Eq.~(\ref{css-gen}), is 
clearly dominated by the all-orders resummed term for $Q_T \le Q$.  At 
larger $Q_T$, the $Y$ term is comparable in size. In the expression for 
$Y$, we use the order-$\alpha_s$ QCD expressions.  Next-to-leading order 
contributions tend to alter theoretical predictions, and, correspondingly, 
we would expect that our predictions would change somewhat for $Q_T > Q$ 
if higher order contributions are included.  

The resummed result (labeled ``total'') makes a smooth transition to the 
fixed-order perturbative result near or just above $Q_T = Q$, for all $Q$, 
without need of a supplementary matching procedure.  With all $Q_T$ 
distributions extended to $Q_T = 100$ GeV, our implementation of the 
resummation formalism works smoothly throughout the $Q_T$ region even for 
the distributions differential in rapidity, $d\sigma/dy dQ_T$.  The 
transition from the resummed distribution to the fixed-order perturbative 
contribution is continuous and less ambiguous than is sometimes seen 
in the literature. 

We note that the resummed results for Higgs boson production begin to 
fall below the fixed-order perturbative expectations when $Q_T$ exceeds 
$Q$.  This effect is likely due to the mismatch at order $\alpha_s$ 
between the resummed term and the asymptotic term.  
A similar effect is not seen in 
our predictions for $Z$ boson production where the fermionic coefficient 
$A_q^{(1)}$ is smaller than the gluonic coefficient $A_g^{(1)}$.  The 
fixed-order result should provide the correct theoretical answer for 
$Q_T \sim Q$ (and above) where only one large momentum scale is present 
and large logarithmic effects proportional to $\ln(Q/Q_T)$ are 
insignificant.  We adopt the fixed-order result as our best estimate of 
the theoretical prediction wherever it is larger than the resummed result 
in the region $Q_T > Q$.   

\begin{figure}[ht]
\centerline{\includegraphics[width=9.0cm]{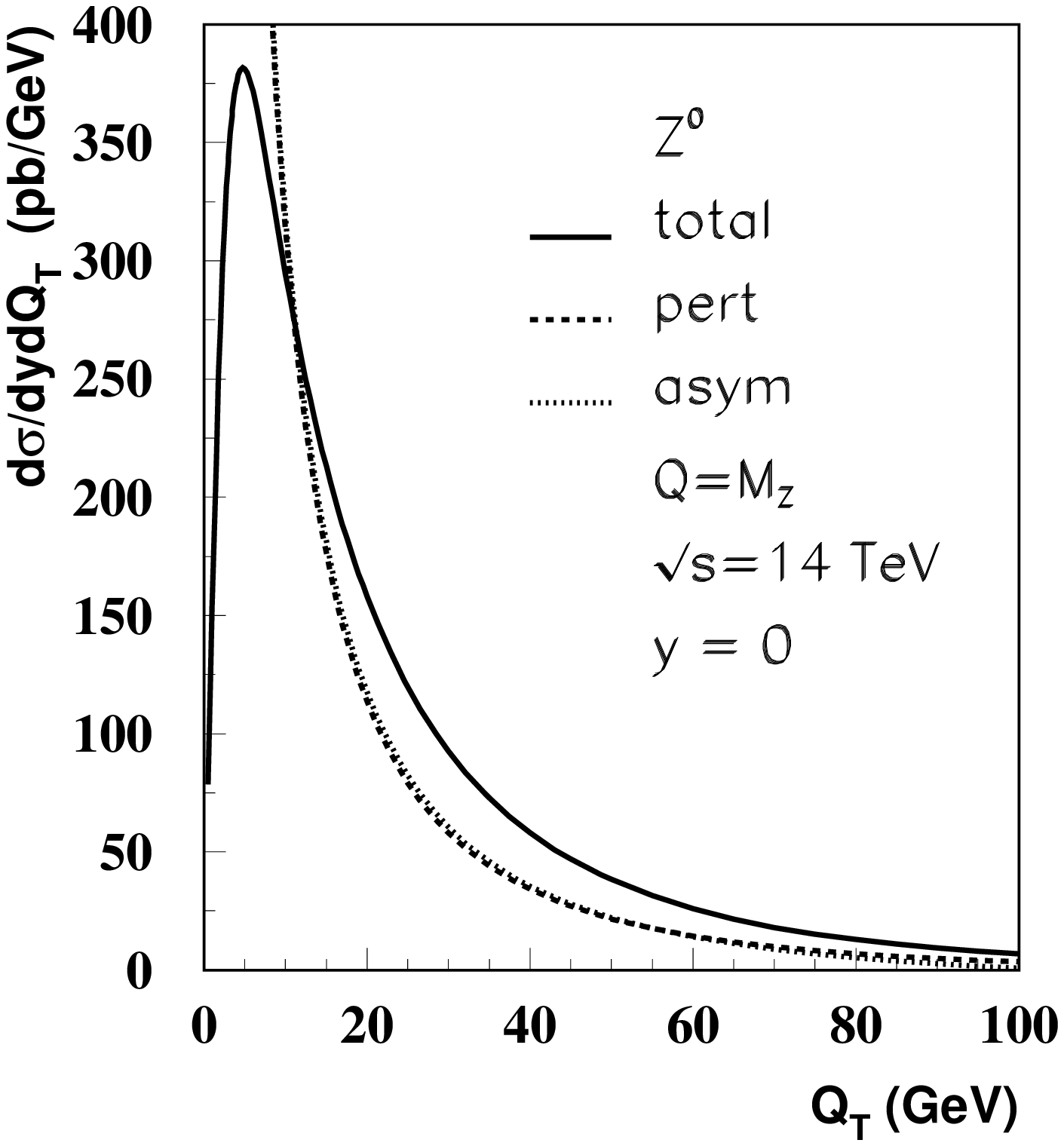}
\includegraphics[width=9.0cm]{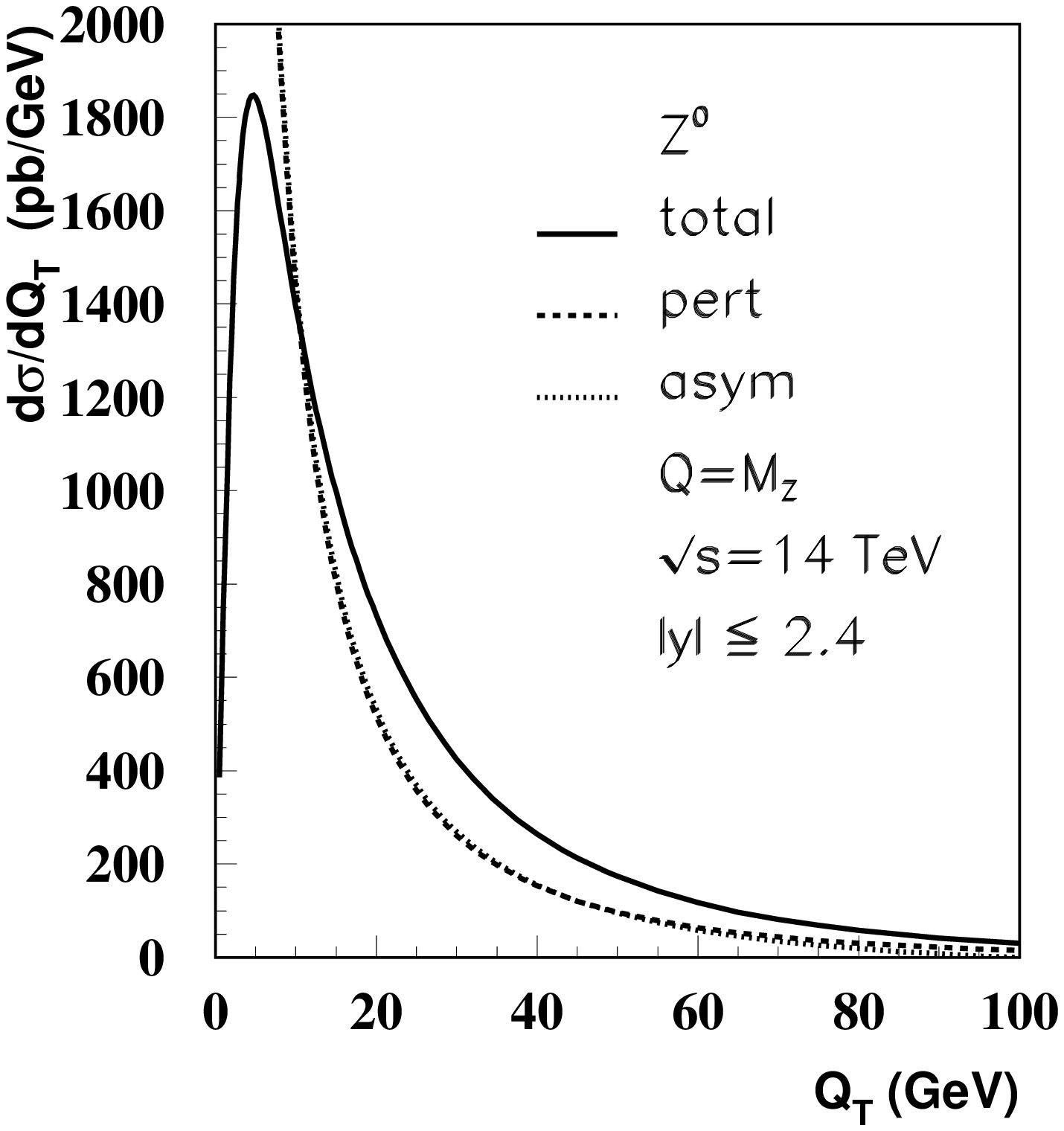}} 
\caption[]{\it Differential cross section $d\sigma/dydQ_T$ 
  as a function of transverse momentum $Q_T$ for $Z$ boson 
  production at $\sqrt{S} = 14$~TeV and (a) fixed rapidity $y=0$ 
  and (b) $|y| < 2.4$. Shown are the total prediction as a 
  solid line, the fixed-order perturbative calculation 
  as a dashed line, and the asymptotic limit as a dotted line.}
\label{fig:QTZ}
\end{figure}

We display our predicted $Q_T$ distributions for $Z$ boson production 
at the LHC in Figs.~\ref{fig:QTZ}.  The resummed distributions are very 
smooth functions of $Q_T$ over the full range shown and they converge 
without discontinuity with the fixed-order perturbative results above 
$Q_T = M_Z$.  No supplementary matching is employed.  

Two points are 
evident in the comparison of $Z$ boson and Higgs boson production, with 
$m_h = M_Z$.  The peak in the $Q_T$ distribution occurs at a 
smaller value of $Q_T$ for $Z$ production.  At $y = 0$, 
the curve peaks at $Q_T \sim 4.8$~GeV for $Z$ production and at 
$Q_T \sim 10$~GeV for Higgs boson production.  Second, the distribution 
is narrower for $Z$ production, falling to half its maximum 
by $Q_T \sim 16$~GeV, whereas the half-maximum for Higgs production is 
not reached until $Q_T \sim 30$~GeV.  Technically, both differences are 
a direct reflection of the differences in the functions $bW(b,Q)$ 
discussed in Sec.~III~A.  The $b$-space distribution is narrower for Higgs 
boson production and peaks at smaller $b$.  Correspondingly, since 
$Q_T \sim 1/b$, the $Q_T$ distribution is broader and peaks at 
larger $Q_T$.  The physics behind this important difference is that the 
larger QCD color factors produce more gluonic showering in the glue-glue 
scattering subprocess that dominates inclusive Higgs boson production than 
in the fermionic subprocesses relevant for $Z$ production.  After all-orders 
resummation, the enhanced showering suppresses the large-$b$ (small $Q_T$) 
region more effectively for Higgs boson production.   

\begin{figure}[ht]
\centerline{\includegraphics[width=9.0cm]{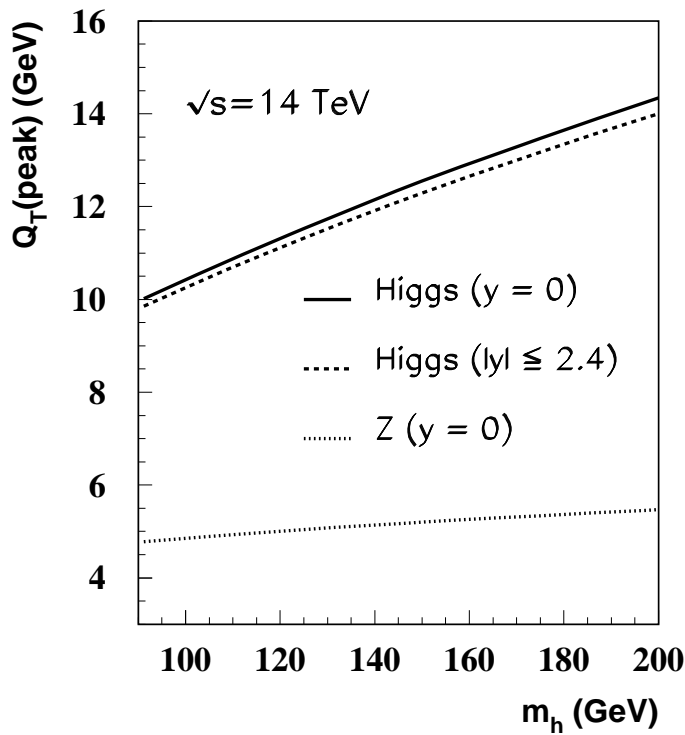}} 
\caption[]{\it Location of the peaks of the transverse momentum 
  distributions for Higgs boson and $Z^*$ boson production as a
  function of Higgs boson mass at $\sqrt{S} = 14$~TeV. For Higgs 
  boson production we show results at fixed rapidity $y=0$ and for 
  $y$ integrated over the interval $|y| \le 2.4$.}
\label{fig:peak}
\end{figure}

Comparison of the predicted $Q_T$ distributions for Higgs boson 
production at different masses of the Higgs boson  
shows that the peak of the distribution shifts to greater 
$Q_T$ as $m_h$ grows and that the distribution broadens somewhat.  
At $y= 0$, the peaks are centered at about $Q_T =$ 10, 11.5, 
12.6, and 14.4 GeV for $m_h = M_Z$, 125, 150, and 200 GeV, 
respectively.  The mass dependence for both Higgs and $Z$ boson 
production is illustrated in Fig.~\ref{fig:peak}.  For $Z^*$ 
boson masses above the physical value $M_Z$ we assume that the 
production model is unchanged except for the difference in mass 
of the $Z^*$.  The peak position at $y=0$ is at 5.5 GeV for 
$M_{Z^*} = 200$~GeV.  
  
\begin{figure}[ht]
\centerline{\includegraphics[width=9.0cm]{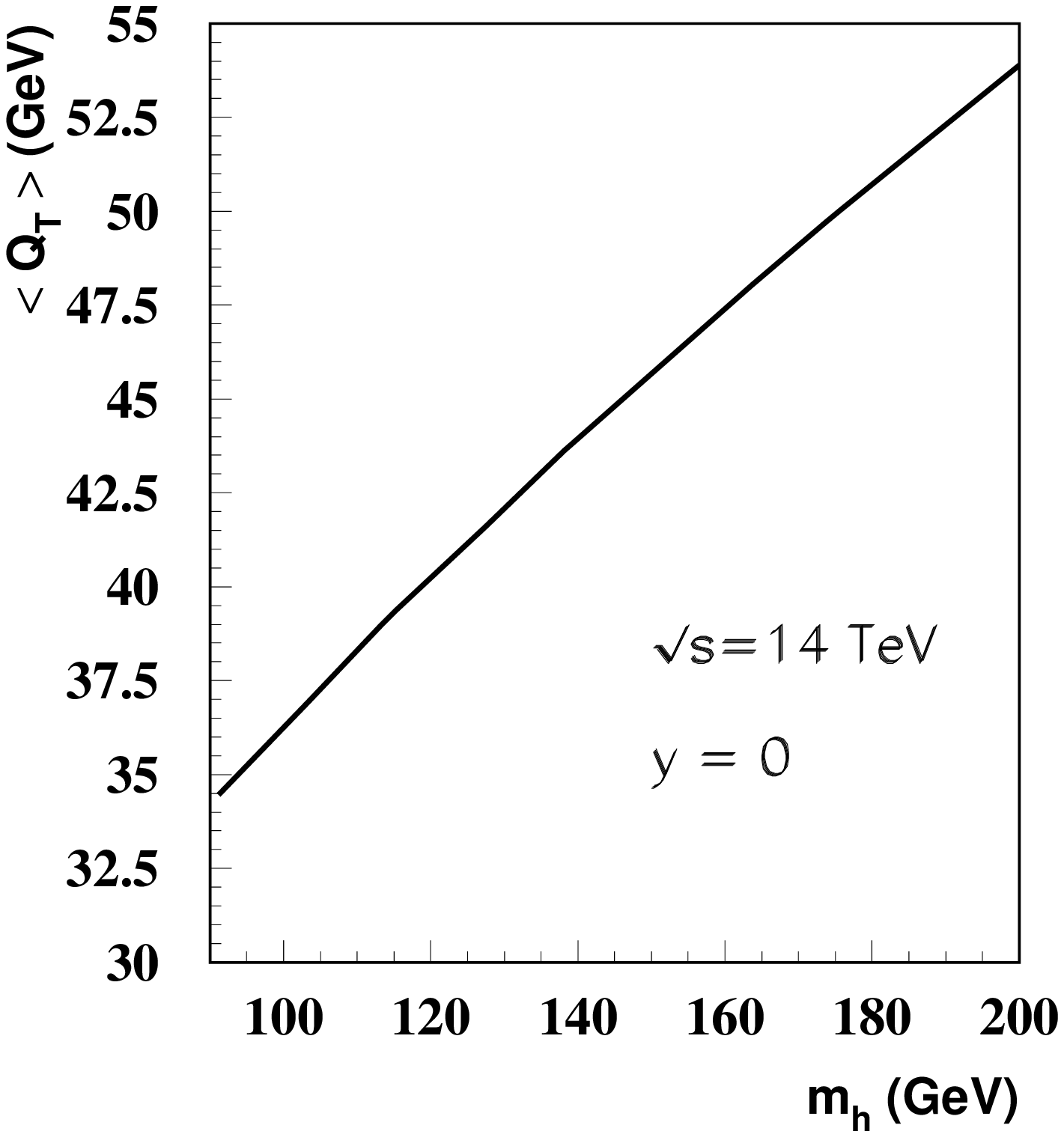}
\includegraphics[width=9.0cm]{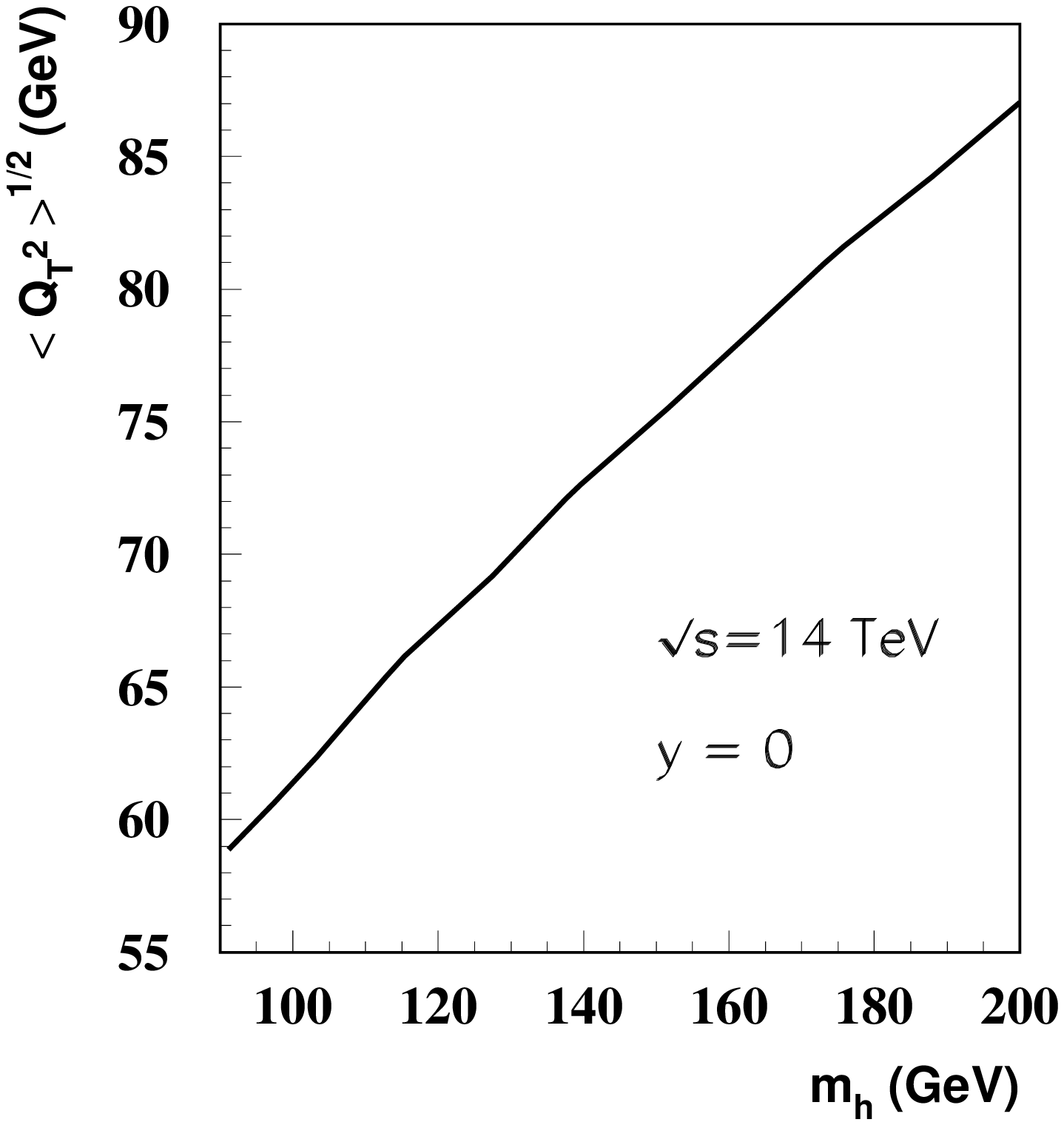}} 
\caption[]{\it Predictions of (a) the average value of the transverse 
  momentum and (b) the root-mean-square for Higgs boson production as a 
  function of Higgs boson mass at $\sqrt{S} = 14$~TeV and fixed rapidity 
  $y=0$.  We show results for Higgs boson masses $m_h$ from $M_Z$ to 
  200 GeV.}
\label{fig:meanQTHiggs}
\end{figure}

The change of the $Q_T$ distribution with $m_h$ can also be 
represented quantitatively with plots of the mean value $<Q_T>$ 
and of the root-mean-square $<Q^2_T>^{\frac{1}{2}}$, shown 
in Fig.~\ref{fig:meanQTHiggs}.  Because of the long tail of the 
distribution $d\sigma/dQ_T$ at large $Q_T$, the mean value $<Q_T>$ 
is about 3.6 times the value of $Q_T$ at which the peak occurs in the 
distribution. We observe that $<Q_T>$ grows from about 35~GeV at 
$m_h = M_Z$ to about 54~GeV at $m_h = 200$ GeV.  
The curve is nearly a straight line over the range shown, with 
$<Q_T> \simeq 0.18 m_h + 18$~GeV.  In computing the averages, we 
switch from the ``total'' result for $d\sigma/dy dQ_T$ shown, for 
example, in Fig.~\ref{fig:QTHiggs0} to the fixed-order perturbative 
result when the fixed-order answer is the larger of the two choices 
above $Q_T = Q$.  We extend our numerical integrations to the 
kinematic limit in the cases of $<Q_T>$ and $<Q^2_T>$ in order to 
achieve insensitivity of the results to 
the upper limit of integration.  The nearly linear growth of $<Q_T>$ 
and of $<Q^2_T>^{1/2}$ with $m_h$ is a reflection of the fact that the 
peak location in $Q_T$ of the distribution $d\sigma/dydQ_T$ grows 
in nearly linear fashion as $m_h$ increases.  Since $Q_T \sim 1/b$, 
the expression for the saddle point, Eq.~(\ref{css-saddle}), 
suggests that $<Q_T> \propto 1/b_{\rm SP} \propto m_h^{\lambda}$, 
with fractional power $\lambda$, rather than a nearly linear 
behavior.  We remark that for large $m_h$, a linear fit is a 
good approximation to the fractional power dependence over a 
restricted range in $m_h$.    

For comparison with our results for Higgs boson production, we quote 
our predictions for $Z$ production: $<Q_T> = 25$~GeV and 
$<Q_T^2>^{1/2} = 38$~GeV.  We note that the difference 
$<Q^h_T> - <Q_T^Z> \simeq 10$ GeV at $m_h = M_Z$, a manifestation of 
more significant gluonic radiation in Higgs boson production.  The 
harder $Q_T$ spectrum suggests that the signal to background ratio 
can be enhanced if Higgs bosons are selected with large $Q_T$, a point 
to which we return in Sec.~VI.   

Choices of variable parameters are made in obtaining our results, 
and it is important to examine the sensitivity of the results to these 
choices.  The parameter choices include the specification of the 
renormalization/factorization scale $\mu$, the non-perturbative 
input, and $Q_{T}^{\rm min}$.  Uncertainties of a different sort are 
associated with the order in perturbation theory at which we work.  
\begin{figure}[ht]
\centerline{\includegraphics[width=9.0cm]{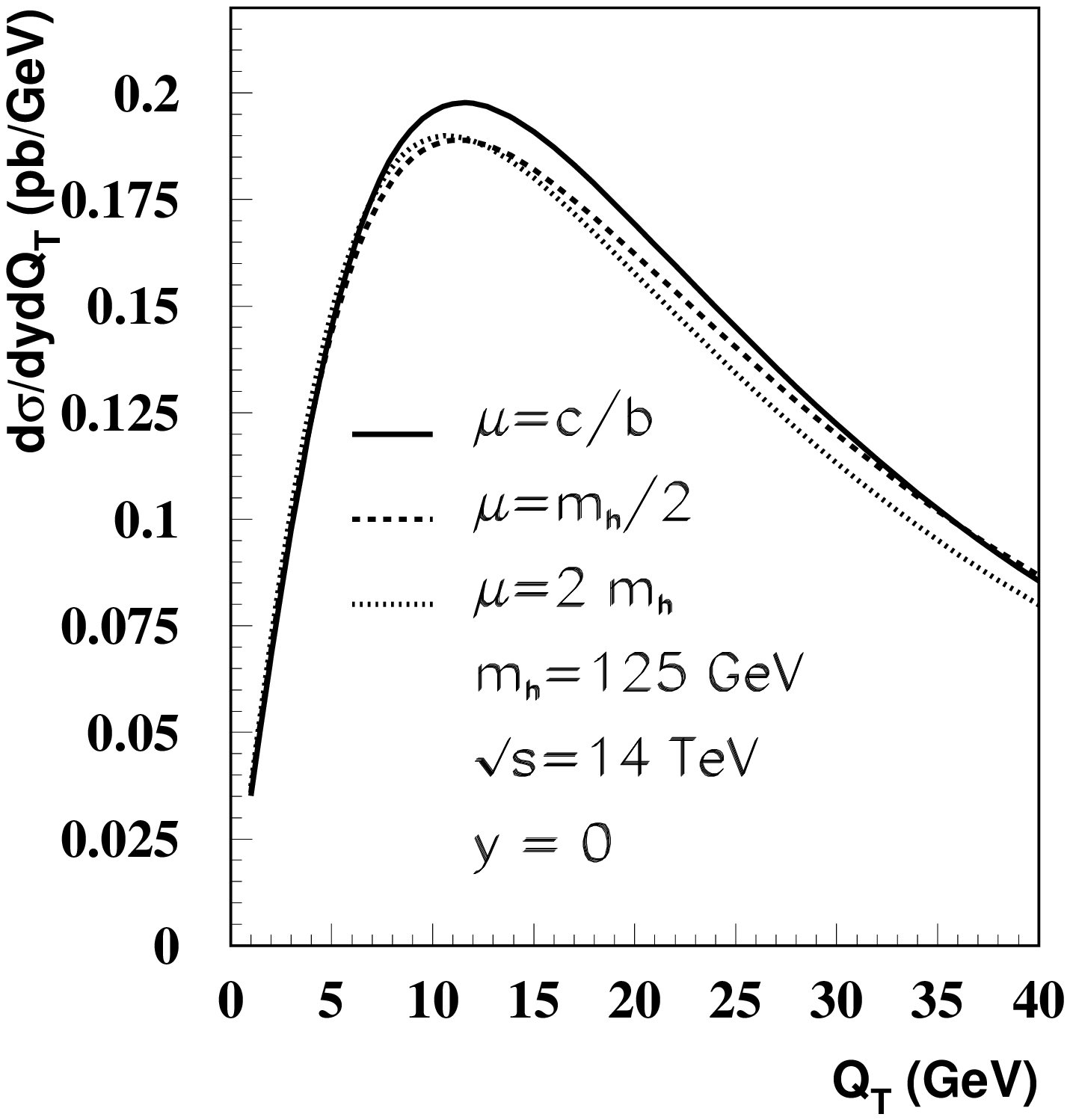}
\includegraphics[width=9.0cm]{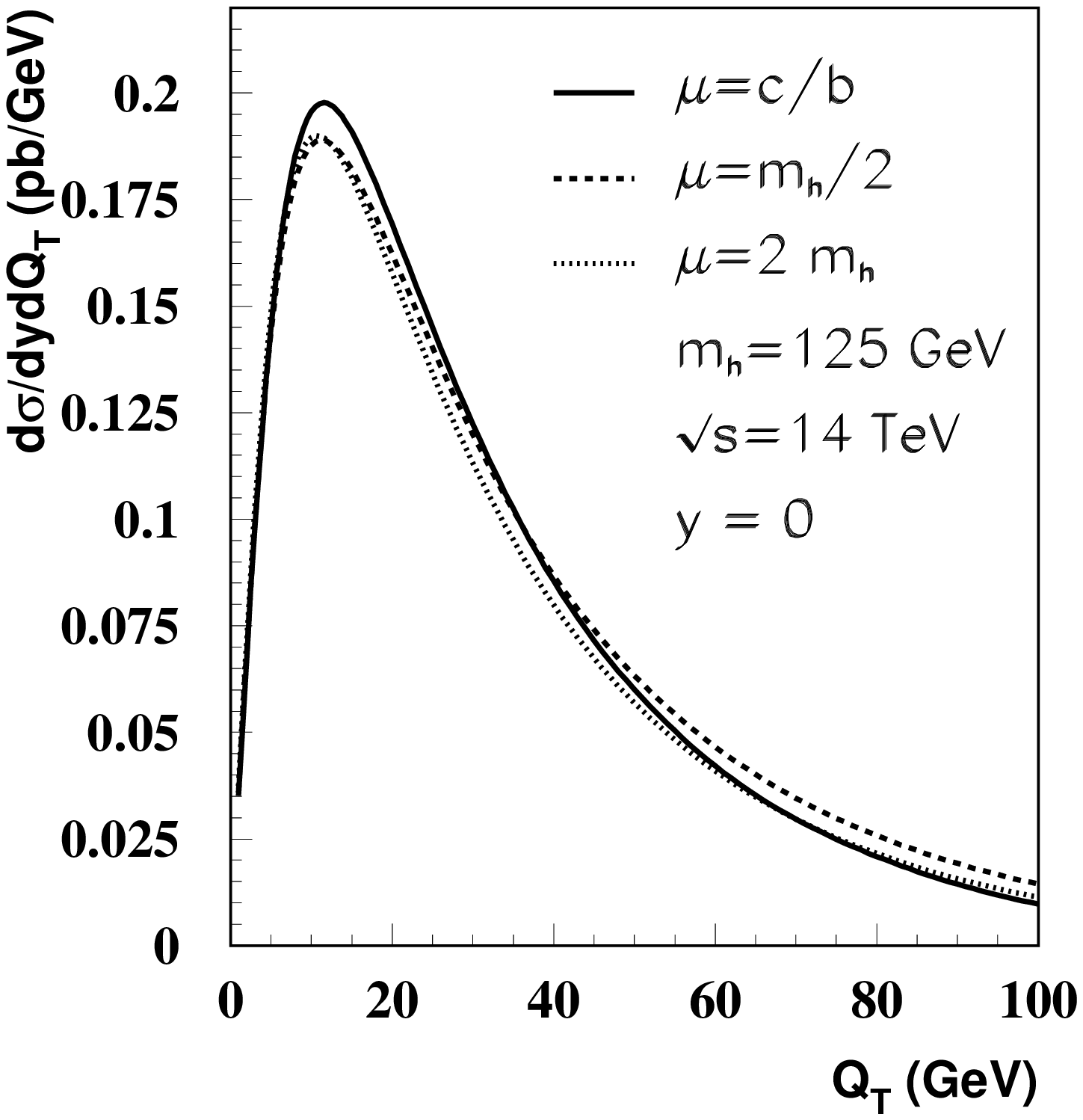}}
\caption[]{\it Renormalization/factorization scale $\mu$ dependence of 
  the differential cross section $d\sigma/dy dQ_T$ as a function of $Q_T$ 
  for Higgs boson mass $m_h = 125$ GeV at $\sqrt{S} = 14$~TeV and fixed 
  rapidity $y=0$.  We show results for three choices of the scale.  In 
  (a) the region of low and intermediate $Q_T$ is shown in 
  expanded form to illustrate the variation of the position in $Q_T$ of 
  the maximum of the distribution.}
\label{fig:scaledep}
\end{figure}

Dependence on the renormalization/factorization scale $\mu$ is often 
a good indicator of theoretical uncertainty.  In Fig.~\ref{fig:scaledep}, 
we show this dependence for the differential cross section $d\sigma/dQ_T$ 
at Higgs boson mass $m_h = 125$ GeV.  Since we are interested principally 
in examining scale dependence of the resummed result, we fix the scale 
in the purely perturbative remainder term Eq.~(\ref{yterm}) to 
be $\mu = 0.5 \sqrt{m_h^2 + Q_T^2}$.  This term is not important in the 
region of modest $Q_T$ where the cross section is large.  We present 
results for three choices of the scale in the resummed term, 
Eqs.~(\ref{css-resum}) and~(\ref{css-pert}).  All these results are 
obtained with $Q_T^{\rm min} = 0.3$~ GeV.  We select as default 
choice $\mu = c/b$ with $c = 2e^{-\gamma_E}$ such that the scale varies 
with the integration variable $b$.  The other two choices are independent 
of $b$ but are proportional to the mass of the Higgs boson, 
$\mu =  0.5 m_h$ and $\mu = 2 m_h$.  Scale dependence is not 
insignificant in the region where the distribution is large. It 
can shift the position of the peak by about 1~GeV, with corresponding 
changes in the normalization of the distribution above and below the 
position of the peak.  The value of $d\sigma/dydQ_T$ at the peak position 
is shifted by 4 to 5\%.  The shift in the location of the peak is about the 
same as the shift that occurs between resummation done at leading-logarithmic 
(LL) accuracy and at NLL accuracy~\cite{Giele:2002hx}.   

\begin{figure}[ht]
\centerline{\includegraphics[width=9.0cm]{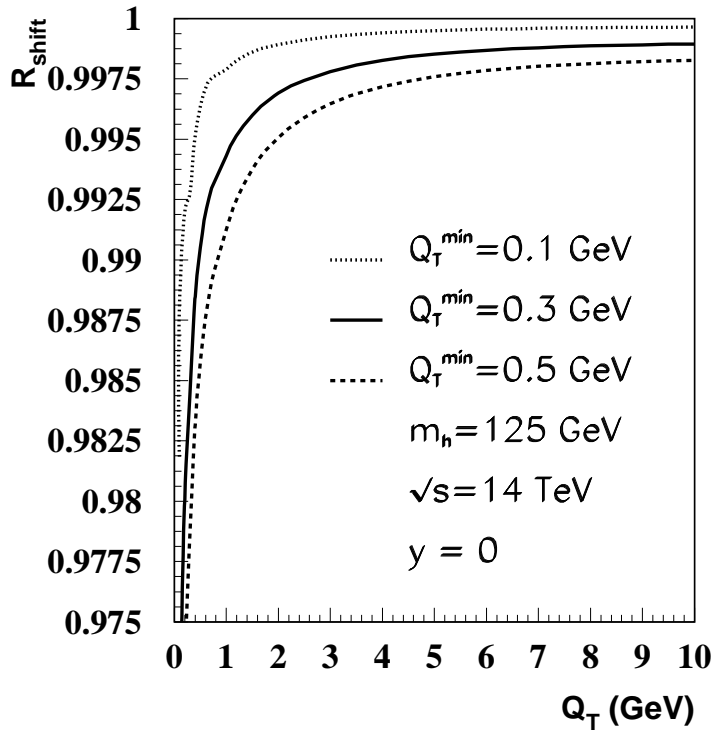}}
\caption[]{\it Dependence of the differential cross section 
  $d\sigma/dy dQ_T$ on the parameter $Q_T^{\rm min}$ as a function 
  of $Q_T$ for Higgs boson mass $m_h = 125$ GeV at $\sqrt{S} = 14$~TeV 
  and fixed rapidity $y=0$.  We show results for three values of 
  $Q_T^{\rm min}$, divided by the results for $Q_T^{\rm min} =0$.}
\label{fig:shiftdep}
\end{figure}
For our 
central results we select $Q_T^{\rm min} = 0.3$~GeV.  We examine changes 
associated with this parameter in Fig.~\ref{fig:shiftdep}.  The ratio 
shown is 
\begin{equation}
R_{\rm shift} = 
{\frac{d\sigma}{dydQ_T} (Q_T^{\rm min})}/\frac{d\sigma}{dydQ_T}(Q_T^{\rm min}=0).
\end{equation}
We vary $Q_T^{\rm min}$ over the range 0 to 500 MeV.  The position of the 
peak of the $Q_T$ distribution is very stable; within a bin width of 
$1$~GeV, there is no change in the location of the peak.  At $m_h = 125$~GeV 
the change in the value of $d\sigma/dydQ_T$ at the peak location of 
$\sim 11.5$~GeV is only 0.2\%  when $Q_T^{\rm min}$ varies from 0 to 0.3 GeV. 
However, for $Q_T < 1$~GeV, the uncertainties are much larger.  One message 
from this exercise is that the resummed $Q_T$ distribution cannot be computed 
reliably when $Q_T \ll Q_T^{\rm peak}$, a conclusion that should not be 
surprising.  On the other hand, the uncertainties in the $Q_T$ distribution in 
the region of small $Q_T$ associated with the choice of $Q_T^{\rm min}$ should 
not be worse than potential uncertainties in the data.  

\begin{figure}[ht]
\centerline{\includegraphics[width=9.0cm]{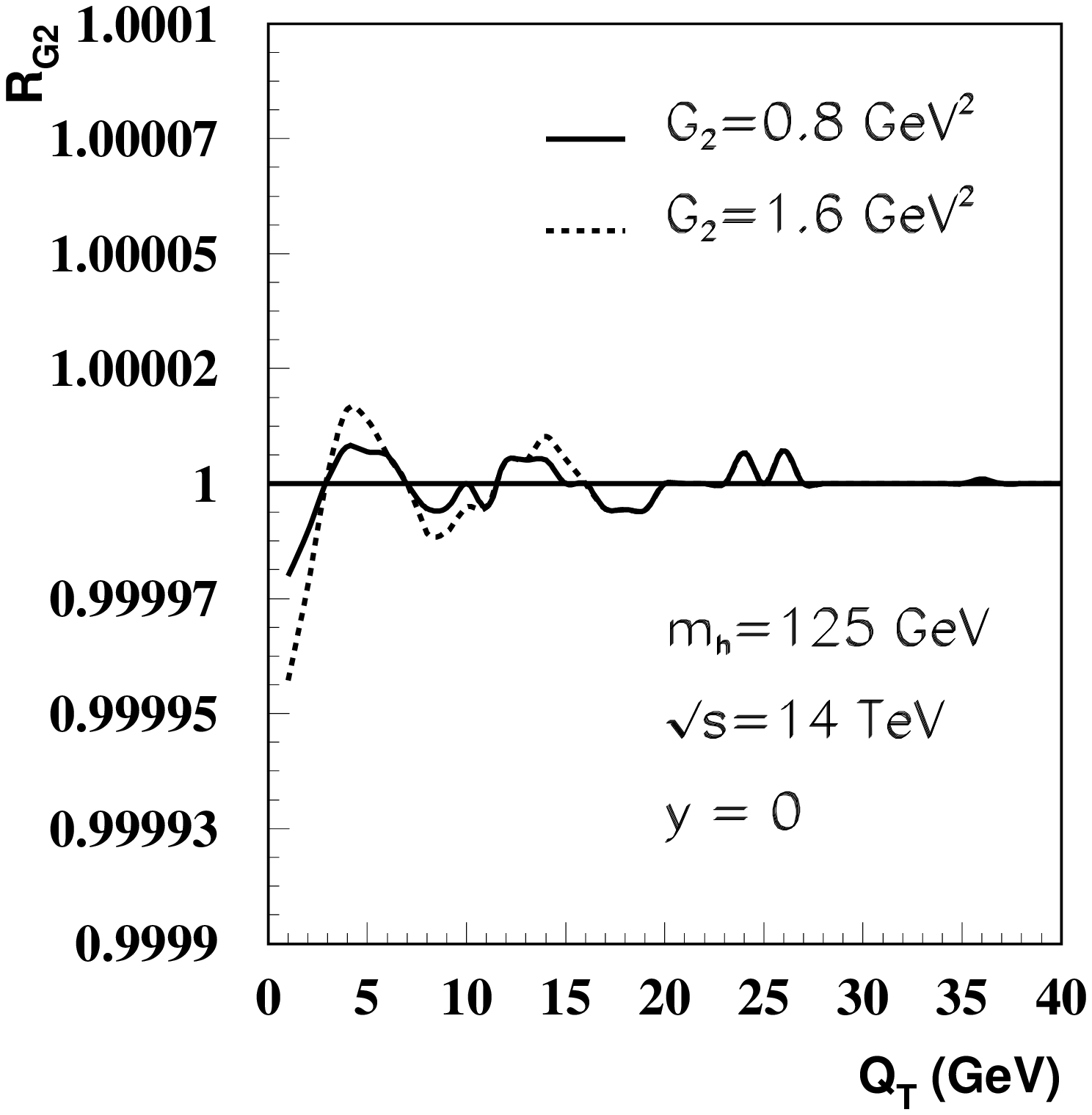}
\includegraphics[width=9.0cm]{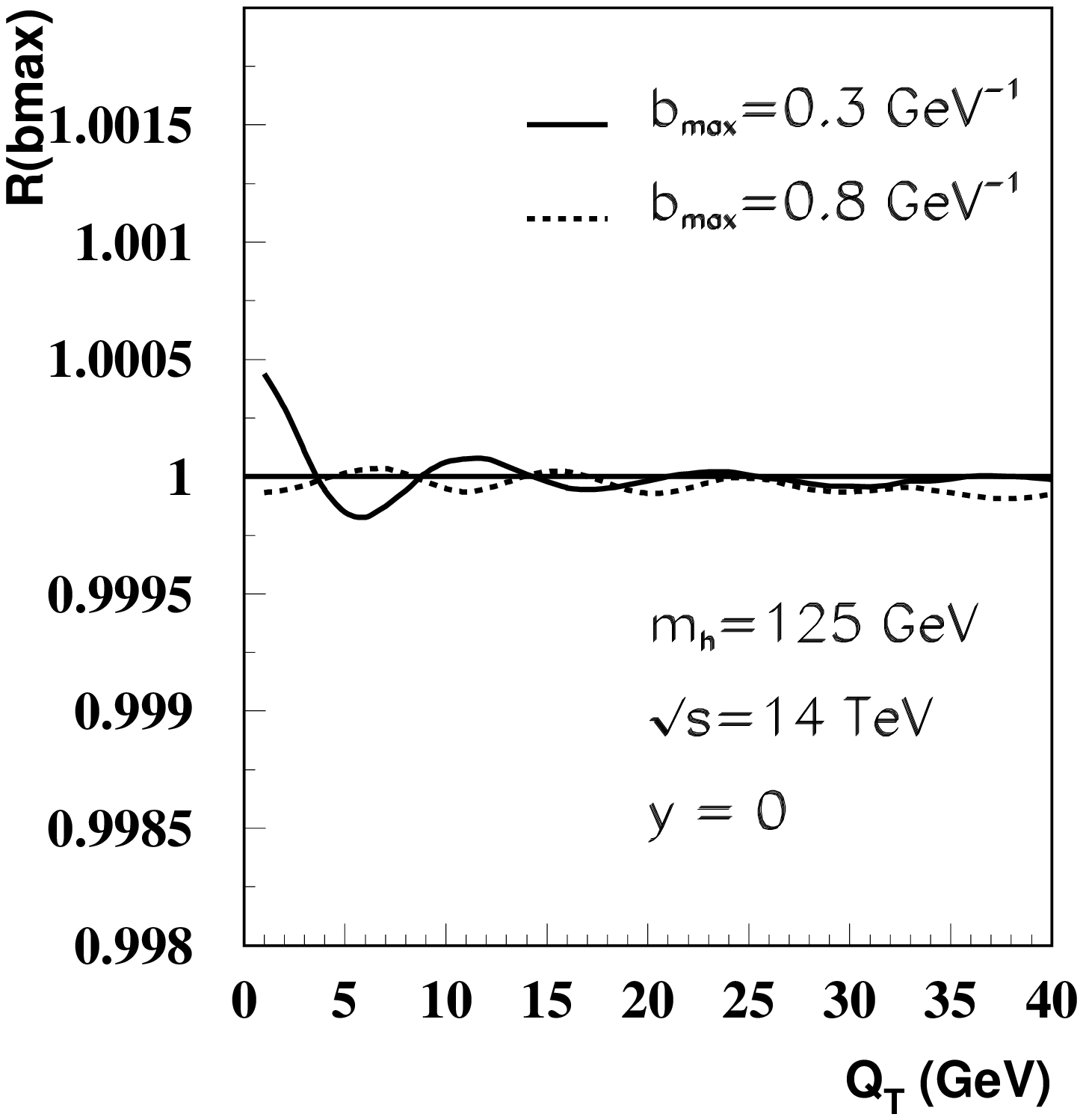}}
\caption[]{\it Dependence of the differential cross section 
  $d\sigma/dy dQ_T$ on the non-perturbative parameters (a) $G_2$ and 
  (b) $b_{\rm max}$ as a function of $Q_T$ 
  for Higgs boson mass $m_h = 125$ GeV at $\sqrt{S} = 14$~TeV and fixed 
  rapidity $y=0$.  We show results for two choices of each parameter.}
\label{fig:nonpertdep}
\end{figure}
In Fig.~\ref{fig:nonpertdep} we examine the dependence of our predictions 
on the choice of the non-perturbative parameters $b_{\rm max}$, $g_2$, and 
$\bar{g}_2$.  We combine the dependence on the latter two parameters in 
a function $G2 = g_2 \ln(Q^2b_{\rm max}^2/c^2) + \bar{g}_2$, defined in 
Ref.~\cite{Qiu:2000hf}.  In Fig.~\ref{fig:nonpertdep}(a) we plot the ratio 
of the $Q_T$ distribution with residual power corrections over its value without 
these contributions.  In Ref.~\cite{Qiu:2000hf}, $G2 = 0.4$~GeV$^2$ is 
shown to produce the best fit to Tevatron data on $W$ and $Z$ boson production.
To enhance the possible influence of power corrections, we choose parameters 
that are a few times larger than are needed in $W$ and $Z$ production.  
In Fig.~\ref{fig:nonpertdep}(a), the values $G2= 0.8$~GeV$^2$ and 
1.6~GeV$^2$ are used, twice- and four-times the values determined earlier.  
The corresponding changes in the predictions for Higgs boson production are 
very small.
In Fig.~\ref{fig:nonpertdep}(b), we examine sensitivities to the choice of 
$b_{\rm max}$.  The ratio displayed is 
$R(b_{\rm max}) = 
d\sigma/dydQ_T(b_{\rm max}) / d\sigma/dydQ_T(b_{\rm max}=0.5~{\rm GeV}^{-1})$.  
For changes in $b_{\rm max}$ between 0.3 and 0.8 GeV$^{-1}$, the uncertainties 
in our predictions are much much less than 1\%.  They are consistent with the 
level of accuracy of our numerical computations.  Evident in 
Fig.~\ref{fig:nonpertdep} is that the formulation we use to describe the
non-perturbative region has no effect on the behavior of differential cross 
section at large $Q_T$.  

Comparison of Figs.~\ref{fig:scaledep} and~\ref{fig:nonpertdep} shows that 
uncertainties related to scale variation are greater than uncertainties 
associated with non-perturbative physics.  Uncertainties associated with 
physics in the region of large $b$ are at most 1 to 2\% depending on the size 
of the power corrections we introduce, whereas scale dependence can 
change distributions by more than 10\%, if one examines the spectrum as a 
whole, and by about 5\% in the region of the peak.  Non-perturbative effects 
vanish very quickly as $Q_T$ increases, while sensitivity to the scale choice 
increases as $Q_T$ increases.  Variations associated with non-perturbative 
physics in the large $b$ region are much smaller at LHC energies than 
higher-order corrections to the perturbative functions $A$, $B$, and 
$C$.  

A different resummation scheme, such as discussed in 
Ref.~\cite{Catani:2000vq}, can lead to uncertainties similar to 
those shown in Fig.~\ref{fig:scaledep} of this paper and in Fig.~7 
of Ref.~\cite{Giele:2002hx}.  Because our method for extrapolation 
into the nonperturbative large $b$ region is different, we expect 
the locations of the maxima in the distributions 
$d\sigma/dy dQ_T$ in Fig.~7 of Ref.~\cite{Giele:2002hx} to shift 
to somewhat larger values of $Q_T$.
\section{CONCLUSIONS}

Discovery of the Higgs particle is essential to shed light on the 
mechanism of electroweak symmetry breaking. The partonic
subprocess $g+g\rightarrow h X$ dominates Higgs boson production in 
hadronic reactions when the Higgs boson mass is in the expected range 
$m_h < 200$~GeV.  The two-scale nature of the production 
dynamics, with mass $m_h$ and transverse momentum $Q_T$ both 
potentially large, and the fact that the fixed-order perturbative QCD 
contributions are singular as $Q_T \rightarrow 0$, necessitates all-orders 
resummation of large logarithmic contributions in order to obtain 
meaningful predictions for the $Q_T$ distribution  
particularly in the region of modest $Q_T$ where the cross section is 
largest.  We perform this resummation of multiple soft-gluon emissions 
using an impact parameter $b$-space formalism.  At LHC energies, the 
typical values of the incident parton momentum fractions 
$x_A\sim x_B\sim m_h/\sqrt{S} \sim 0.009$ (for $m_h = 125$~GeV) are 
small, and the gluon distribution evolves steeply at small $x$. 
Consequently, the saddle point in $b$ of the Fourier transform 
from $b$-space to $Q_T$ space is well into the region of perturbative 
validity.  The resummed $Q_T$ distributions are therefore determined 
primarily by the perturbatively calculated $b$-space distributions at 
small $b$, with negligible contributions from the 
non-perturbative region of large $b$. Our resummation
formalism has excellent predictive power for the $Q_T$ distributions of 
Higgs boson and $Z$ production at the LHC energy, $\sqrt{S}=14$~TeV,  
for $Q_T$ as large as the respective boson masses $Q$.  
An examination of the numerical sensitivity of 
our results to variations in the renormalization/factorization scale 
$\mu$ and to parameters associated with the form of the assumed 
non-perturbative function at large $b$ shows the expected modest 
dependence of $\mu$ and essentially no dependence on the non-perturbative 
parameters, except in the region of extremely small $Q_T$.   

In this paper, we present predictions for the $Q_T$ distributions of Higgs 
boson and $Z$ production at $\sqrt{S}=14$~TeV.  Results are shown in 
Figs.~\ref{fig:QTHiggs0} - \ref{fig:meanQTHiggs}.  At the same mass, 
$m_h = M_Z$, the predicted mean value $<Q_T>$ is about 10~GeV larger for 
Higgs boson production than for $Z$ boson production.  For the Higgs boson, 
$<Q_T>$ grows from about 35~GeV at $m_h = M_Z$ to about 54~GeV at 
$m_h = 200$ GeV, and the root-mean-square $<Q_T^2>^{1/2}$ from about 59~GeV 
to about 87~GeV.   

Searches for the Higgs boson in its decay to two photons require an 
excellent understanding of the production characteristics of 
both the signal and backgrounds.  The primary background to direct 
photon production comes from decays of hadrons such as $\pi^0$, 
$\eta^0$, and $\omega^0$, themselves produced copiously at large 
transverse momentum from the fragmentation of jets, as well as from 
misidentified photons ({\em e.g.}, from processes that produce two jets 
or a jet and a photon and in which jets are misidentified as photons).  
Even after excellent photon/jet discrimination and application of 
photon isolation restrictions, simulations show ratios of the expected 
signal to irreducible backgrounds in the range of 4 to 
5~$ \times 10^{-2}$ for $m_h$ in the interval 120 to 
130 GeV~\cite{atlas:1999fr}. The irreducible backgrounds arise from 
the QCD annihilation $q \bar{q} \rightarrow \gamma \gamma$, the Compton 
$q g \rightarrow \gamma \gamma q$ (with one of the final photons 
produced from parton fragmentation (generalized bremsstrahlung)), and 
the gluon box $g g \rightarrow \gamma \gamma$ subprocesses, and their 
counterparts at higher orders of perturbation 
theory~\cite{Berger:1983yi,Balazs:1997hv,Bern:2002jx}.  Although there 
is some dependence on the di-photon invariant mass and photon 
transverse momentum, the contributions of the annihilation and gluon 
box processes are comparable, and the fragmentation processes make 
a smaller contribution than either of these others.  Of interest to 
us is the expected dependence of the ratio of signal to irreducible 
background.  In this paper we provide predictions for the transverse 
momentum distribution of the signal.  The gluon box subprocess 
has the same initial state as the inclusive gluon fusion process that 
produces the Higgs boson.  Consequently, for di-photon invariant mass 
$M_{\gamma \gamma} = m_h$, we expect that the transverse momentum 
spectrum of the gluon box contribution to the irreducible background will 
have the same shape after soft gluon resummation as that for the Higgs 
boson.  However, the other major component of the irreducible background, 
the annihilation $q \bar{q} \rightarrow \gamma \gamma$ subprocess, has 
the same initial state structure as $Z$ boson production.  As we show in 
this paper, the $Q_T$ spectrum of $Z$ production is predicted to be softer 
than that for Higgs boson production ({\em c.f.}, 
Figs.~\ref{fig:peak} and \ref{fig:meanQTHiggs}).       
We suggest therefore that a selection of events with large 
$Q_T^{\gamma \gamma}$ could help in significantly improving the 
signal to background ratio.   
\section*{ACKNOWLEDGMENTS}

Research in the High Energy Physics Division at Argonne is supported 
by the United States Department of Energy, Division of High Energy 
Physics, under Contract W-31-109-ENG-38.  JWQ is supported in part by 
the United States Department of Energy under Grant No. DE-FG02-87ER40371.  
JWQ acknowledges the hospitality of the Argonne high energy physics 
theory group while part of this research was being carried out.  ELB 
acknowledges the hospitality of the Aspen Center for Physics where 
part of this work was done.

\end{document}